\documentclass[12pt]{article}

\usepackage{amsmath}
\usepackage{amsfonts}
\usepackage{amscd}
\usepackage{amsthm}
\usepackage{setspace}
\usepackage{graphicx}
\usepackage{authblk}
\usepackage{caption}
\usepackage{ytableau}
\usepackage{mathtools}
\usepackage[all,cmtip]{xy}
\usepackage[numbers,compress]{natbib}

\def\Z{\mathbb{Z}}

\def\C{\mathbb{C}}
\def\P{\mathbb{P}}

\def\til{\tilde}

\begin{document}

\begin{titlepage}

\begin{flushright}
YITP-16-87
\end{flushright}

\vskip 1cm

\begin{center}

{\large Gauge groups and matter spectra in F-theory compactifications on genus-one fibered Calabi-Yau 4-folds without section - hypersurface and double cover constructions}

\vskip 1.2cm

Yusuke Kimura$^1$
\vskip 0.4cm
{\it $^1$Yukawa Institute for Theoretical Physics, Kyoto University, Kyoto 606-8502, Japan}
\vskip 0.4cm
E-mail: kimura@yukawa.kyoto-u.ac.jp

\vskip 1.5cm
\abstract{\par We investigate gauge theories and matter contents in F-theory compactifications on families of genus-one fibered Calabi--Yau 4-folds lacking a global section. To construct families of genus-one fibered Calabi--Yau 4-folds that lack a global section, we consider two constructions: hypersurfaces in a product of projective spaces, and double covers of a product of projective spaces. We consider specific forms of defining equations for these genus-one fibrations, so that genus-one fibers possess complex multiplications of specific orders. These symmetries enable a detailed analysis of gauge theories. $E_6$, $E_7$, and $SU(5)$ gauge groups arise in some models. Discriminant components intersect with one another in the constructed models, and therefore, discriminant components contain matter curves. We deduce potential matter spectra and Yukawa couplings.}  

\end{center}
\end{titlepage}

\tableofcontents
\section{Introduction}
F-theory \cite{Vaf, MV1, MV2} is a framework that extends the type IIB superstring theory to a nonperturbative regime, and the compactification geometries for F-theory are Calabi--Yau manifolds with a torus fibration. In the F-theory approach, the modular parameter of a genus-one curve, as a fiber of a torus fibration, is identified with the axio-dilaton; this formulation enables the axio-dilaton to have $SL_2(\Z)$ monodromy. Local F-theory models have been mainly discussed in recent studies on F-theory model building \cite{DWmodel, BHV1, BHV2, DW}. However, to deal with the issues of gravity and the early universe including inflation, global geometries of F-theory compactifications need to be considered. We investigate the geometries of F-theory compactifications from the global perspective in this study.
\par A Calabi--Yau manifold with a torus fibration may or may not admit a global section. F-theory models on Calabi--Yau manifolds with a global section have been studied previously, for example, in \cite{GW, KMSS, GKPW, MP, MPW, BGK, BMPWsection, CKP, CGKP, CKPaddendum, CKPS, BMPW, GK, CKPT, CGKPS}. In recent years, there has been an increasing interest in F-theory models on Calabi--Yau genus-one fibrations without a global section\footnote{\cite{BEFNQ, BDHKMMS} discussed F-theory compactifications without a global section.}. Initiated in \cite{BM, MTsection}, F-theory compactifications lacking a global section have been discussed in recent studies. See also, for example, \cite{AGGK, KMOPR, GGK, MPTW, MPTW2, BGKintfiber, CDKPP, LMTW, GKK, K, K2} for recent advances in F-theory models that lack a global section. It was argued in \cite{MTsection} that, by considering the Jacobian fibrations, the F-theory models on Calabi--Yau genus-one fibrations without a global section can be related to the geometry of Calabi--Yau elliptic fibrations with a section. 
\par In this note, we construct genus-one fibered Calabi--Yau 4-folds without a global section, and we use these spaces as compactification geometries for F-theory to investigate F-theory models without a section. We consider two constructions: hypersurfaces in a product of projective spaces, and double covers of a product of projective spaces, to construct genus-one fibered Calabi--Yau 4-folds without a rational section. In these constructions, we consider Calabi--Yau 4-folds whose discriminant components intersect with one another. Therefore, a component contains matter curves. Matter\footnote{See, for example, \cite{BIKMSV, KV, GM, MTmatter, GM2} for the correspondence of the singularities of Calabi--Yau manifolds and the associated matter contents. Matter arising from the structure of divisor is discussed in \cite{KMP, Pha}. For discussion of the deformation and the resolution of singularities of manifolds, see, for example, \cite{KM}. For analysis of matter in four-dimensional (4d) F-theory with flux, see, e.g., \cite{DWmodel, BHV1}.} with non-trivial chirality arises in F-theory models considered in this note. We discuss gauge theories and matter contents in F-theory compactified on such Calabi--Yau 4-folds. In the two constructions of genus-one fibered Calabi--Yau 4-folds without a section, we particularly focus on the families given by specific equations. The specific equations of genus-one fibered Calabi--Yau 4-folds that we choose enable a detailed investigation of the gauge theories in F-theory models.
\par In this note, we take a direct approach to deduce physical information directly from the defining equations of the constructed genus-one fibered Calabi--Yau 4-folds without a section. We consider two families of hypersurfaces in a product of projective spaces, which we refer to as ``Fermat-type hypersurfaces'' and ``hypersurfaces in Hesse form''\footnote{Similar conventions of terms were made for K3 hypersurfaces in \cite{K}.}; one family of double covers of a product of projective spaces given by equations of a specific form. Among the families of genus-one fibered Calabi--Yau 4-folds without a global section that we consider in this study, genus-one fibers of Fermat-type hypersurfaces and double covers of a product of projective spaces (given by equations of specific forms) possess particular symmetries; these symmetries of genus-one fibers strictly limit possible monodromies around the singular fibers. Consequently, these symmetries greatly constrain possible non-Abelian gauge groups that can form on the 7-branes. We deduce the non-Abelian gauge symmetries arising on the 7-branes in F-theory models, and utilizing these constraints imposed by the symmetries of genus-one fibers, we perform a consistency check of our results. \footnote{Similar consistency checks of non-Abelian gauge symmetries that form on the 7-branes can be found in \cite{K, K2}.} 
\par Concretely, we consider multidegree (3,2,2,2) hypersurfaces in $\P^2\times\P^1\times\P^1\times\P^1$, and double covers of $\P^1\times\P^1\times\P^1\times\P^1$ ramified over a multidegree (4,4,4,4) 3-fold. We find that, in F-theory compactifications on Fermat-type (3,2,2,2) hypersurfaces, generically $SU(3)$ gauge symmetries arise on the 7-branes, and when the 7-branes coincide, $SU(3)$ symmetries on the 7-branes collide and are enhanced to $E_6$ symmetry. Only gauge symmetries of type $SU(N)$ arise on the 7-branes in F-theory compactifications on (3,2,2,2) hypersurfaces in Hesse form. In F-theory compactifications on double covers of $\P^1\times\P^1\times\P^1\times\P^1$ ramified over a multidegree (4,4,4,4) 3-fold (given by equations of specific form), generically $SU(2)$ gauge symmetries arise on the 7-branes. When the 7-branes coincide, $SU(2)$ gauge symmetries collide and are enhanced to $SO(7)$ symmetry; when more 7-branes coincide, gauge symmetries are enhanced further to $E_7$ symmetry. 
\par We compute the Jacobian fibrations of the families of genus-one fibered Calabi--Yau 4-folds without a global section. We determine the Mordell--Weil groups of the Jacobian fibrations of specific members of the family of Fermat-type hypersurfaces, and the family of double covers. In particular, we deduce that F-theory compactifications on these specific members do not have a $U(1)$ gauge symmetry. 
\par We also discuss potential matter contents and potential Yukawa couplings. As will be discussed in Section \ref{sec 4}, when we consider algebraic 2-cycles as candidates for four-form fluxes \footnote{Four-form flux and a generated superpotential were studied in \cite{GVW}. See, for example, \cite{MSSN, CS, MSSN2, BCV, MS, KMW, GH, KMW2, IJMMP, KSN, CGK, BCV2, BKL, LMTW, SNW, LW} for recent progress of four-form flux in F-theory.}, we need to consider {\it intrinsic} algebraic 2-cycles \footnote{We explain what we mean by the term ``intrinsic algebraic cycles'' in Section \ref{ssec 4.2}.}. We need to compute their self-intersections to see if they can cancel the tadpole; however, it is technically difficult to compute the self-intersection of an intrinsic algebraic 2-cycle in the geometry of Calabi--Yau 4-folds that we consider in this note. We only deduce the potential matter contents, and potential Yukawa couplings. We compute the Euler characteristics of the constructed Calabi--Yau 4-folds, to derive constraints imposed on the self-intersection of a four-form flux to cancel the tadpole. 
\par The outline of this note is as follows: In Section \ref{sec 2}, we introduce the two constructions of genus-one fibered Calabi--Yau 4-folds without a section. The constructions use hypersurfaces in a product of projective spaces, and double covers of a product of projective spaces; to perform a detailed study of gauge theories, we only consider families given by specific equations in these constructions. We determine the discriminant loci and their components. We describe the forms of the discriminant components. In Section \ref{sec 3}, we deduce the non-Abelian gauge symmetries arising on the 7-branes in F-theory compactifications on the families of genus-one fibered Calabi--Yau 4-folds lacking a global section, as introduced in Section \ref{sec 2}. We choose the defining equations of Fermat-type Calabi--Yau hypersurfaces, and Calabi--Yau 4-folds constructed as double covers, so that genus-one fibers possess complex multiplications of specific orders. These particular symmetries constrain possible non-Abelian gauge groups that can form on 7-branes. We confirm that the non-Abelian gauge groups that we deduce are in agreement with these constraints. This gives a consistency check of our solutions. In Section \ref{sec 4}, we consider the existence of a consistent four-form flux. We compute the Euler characteristics of Calabi--Yau 4-folds, to derive conditions for the self-intersections of four-form fluxes to cancel the tadpole. In Section \ref{sec 5}, we determine the potential matter spectra, and potential Yukawa couplings. In Section \ref{sec 6}, we state our conclusions. 

\section{Genus-One Fibered Calabi--Yau 4-folds without a Global Section, and Discriminant Loci}
\label{sec 2}
In this section, we construct genus-one fibered Calabi--Yau 4-folds that lack a global section. We consider the following two constructions: 
\begin{itemize}
\item multidegree (3,2,2,2) hypersurfaces in $\P^2\times\P^1\times\P^1\times\P^1$
\item double covers of $\P^1\times\P^1\times\P^1\times\P^1$ branched along a multidegree (4,4,4,4) 3-fold.
\end{itemize}
These two constructions have the trivial canonical bundles $K=0$, and they are therefore Calabi--Yau 4-folds. Furthermore, natural projections onto $\P^1\times\P^1\times\P^1$ give genus-one fibrations, so they are genus-one fibered. Additionally, they have natural projections onto $\P^1\times\P^1$, which give K3 fibrations. 
\par For each of these two constructions, we only consider families given by specific equations, whose symmetries allow for a detailed investigation of gauge theories. Gauge theories in F-theory on the families of Calabi--Yau 4-folds will be discussed in Section \ref{sec 3}. In this section, we introduce the families of genus-one fibered Calabi--Yau 4-folds given by specific equations. We show that they do not admit a global section. We determine the discriminant loci of the families of Calabi--Yau 4-folds, and we describe the forms of the discriminant components. 

\subsection{Multidegree (3,2,2,2) Hypersurfaces in $\P^2\times\P^1\times\P^1\times\P^1$}
\subsubsection{Two Types of Equations for (3,2,2,2) Hypersurfaces}
\label{sssec 2.1.1}
Multidegree (3,2,2,2) hypersurfaces in $\P^2\times\P^1\times\P^1\times\P^1$ are Calabi--Yau 4-folds. A fiber of the natural projection onto $\P^1\times\P^1\times\P^1$ is a degree 3 hypersurface in $\P^2$, which is a genus-one curve; therefore, (3,2,2,2) hypersurfaces in $\P^2\times\P^1\times\P^1\times\P^1$ are genus-one fibration over the base 3-fold $\P^1\times\P^1\times\P^1$. A fiber of a natural projection onto $\P^1\times\P^1$ is a bidegree (3,2) hypersurface in $\P^2\times\P^1$, which is a genus-one fibered K3 surface, and therefore, projection onto $\P^1\times\P^1$ gives a K3 fibration. 
\par In this note, we particularly focus on two families of (3,2,2,2) hypersurfaces given by the following two types of equations:
\begin{equation}
(t-\alpha_1)(t-\alpha_2)fX^3+(t-\alpha_3)(t-\alpha_4)gY^3+(t-\alpha_5)(t-\alpha_6)hZ^3=0
\label{hypersurf 1 ; 2.1.1}
\end{equation}
\begin{equation}
\begin{split}
(t-\beta_1)(t-\beta_2)aX^3+(t-\beta_3)(t-\beta_4)bY^3+(t-\beta_5)(t-\beta_6)cZ^3 & \\
-3(t-\beta_7)(t-\beta_8)dXYZ & =0.
\end{split}
\label{hypersurf 2 ; 2.1.1}
\end{equation}
$[X:Y:Z]$ is homogeneous coordinates on $\P^2$, and $t$ is the inhomogeneous coordinate on the first $\P^1$ in $\P^2\times \P^1\times \P^1 \times \P^1$. $\alpha_i$ ($i=1,\cdots,6$) and $\beta_j$ ($j=1,\cdots,8$) are points in this first $\P^1$. $f,g,h$ and $a,b,c,d$ are bidegree (2,2) polynomials on $\P^1\times\P^1$, where the $\P^1$'s in the product $\P^1\times\P^1$ are the last two $\P^1$'s in $\P^2\times \P^1\times \P^1 \times \P^1$. 
\par We refer to the family of hypersurfaces given by the first type equation (\ref{hypersurf 1 ; 2.1.1}) as {\it Fermat-type} hypersurfaces, and we refer to the family of hypersurfaces given by the second type equation (\ref{hypersurf 2 ; 2.1.1}) as hypersurfaces in {\it Hesse form}. 
\par For Fermat-type hypersurface (\ref{hypersurf 1 ; 2.1.1}), a K3 fiber of the projection onto the product $\P^1\times\P^1$ of the second and third $\P^1$'s is described by the following equation:
\begin{equation}
(t-\alpha_1)(t-\alpha_2)\, X^3+(t-\alpha_3)(t-\alpha_4)\, Y^3+(t-\alpha_5)(t-\alpha_6)\, Z^3=0
\label{K3 1 ; 2.1.1}
\end{equation}
This is Fermat-type K3 hypersurface, which is discussed in \cite{K}. Similarly, for the hypersurface in Hesse form (\ref{hypersurf 2 ; 2.1.1}), a K3 fiber of the projection onto the product $\P^1\times\P^1$ of the second and third $\P^1$'s is given by the following equation:
\begin{equation}
\begin{split}
(t-\beta_1)(t-\beta_2)\, X^3+(t-\beta_3)(t-\beta_4)\, Y^3+(t-\beta_5)(t-\beta_6)\, Z^3 & \\
-3(t-\beta_7)(t-\beta_8)\, XYZ & =0.
\end{split}
\label{K3 2 ; 2.1.1}
\end{equation}
This is K3 hypersurface in Hesse form, which is discussed in \cite{K}.
\par In \cite{K}, it was shown that Fermat-type K3 hypersurfaces (\ref{K3 1 ; 2.1.1}) and K3 hypersurfaces in Hesse form (\ref{K3 2 ; 2.1.1}) are genus-one fibered, but their generic members lack a global section to the fibration. If Fermat-type (3,2,2,2) Calabi--Yau hypersurfaces (\ref{hypersurf 1 ; 2.1.1}) admit a rational section, it restricts as a global section to the K3 fiber. This means that Fermat-type K3 hypersurfaces (\ref{K3 1 ; 2.1.1}) admit a global section, which is a contradiction. Similar reasoning applies to (3,2,2,2) Calabi--Yau hypersurfaces in Hesse form (\ref{hypersurf 2 ; 2.1.1}). We therefore conclude that Fermat-type (3,2,2,2) Calabi--Yau hypersurfaces (\ref{hypersurf 1 ; 2.1.1}) and Calabi--Yau hypersurfaces in Hesse form (\ref{hypersurf 2 ; 2.1.1}) are genus-one fibered, but they lack a rational section. 

\subsubsection{Discriminant Locus and Forms of Discriminant Components of Fermat-type (3,2,2,2) Hypersurfaces}
\label{sssec 2.1.2}
We determine the discriminant locus, and the forms of the discriminant components of Fermat-type (3,2,2,2) hypersurface
\begin{equation}
(t-\alpha_1)(t-\alpha_2)fX^3+(t-\alpha_3)(t-\alpha_4)gY^3+(t-\alpha_5)(t-\alpha_6)hZ^3=0.
\label{hypersurf ; 2.1.2}
\end{equation}
A genus-one fibered Calabi--Yau 4-fold and its Jacobian fibration have identical discriminant loci. We deduce the discriminant components of Fermat-type (3,2,2,2) Calabi--Yau hypersurface (\ref{hypersurf ; 2.1.2}) by studying the Jacobian fibration.
\par The Jacobian fibration of Fermat-type hypersurface (\ref{hypersurf ; 2.1.2}) is given by the following equation:
\begin{equation}
\label{jacobianFermat ; 2.1.2}
X^3+Y^3+\Pi_{i=1}^6 (t-\alpha_i)\cdot fgh\cdot Z^3=0.
\end{equation}  
The Jacobian fibration (\ref{jacobianFermat ; 2.1.2}) transforms into the following Weierstrass form \cite{Cas} 
\begin{equation}
\label{localjacobian ; 2.1.2}
y^2=x^3-2^4\cdot 3^3\cdot \Pi_{i=1}^6 (t-\alpha_i)^2\cdot f^2g^2h^2. 
\end{equation}
Therefore, the discriminant of the Jacobian fibration (\ref{jacobianFermat ; 2.1.2}) is given by the following equation:
\begin{equation}
\Delta \sim \Pi_{i=1}^6 (t-\alpha_i)^4\cdot f^4g^4h^4.
\end{equation} 
The discriminant locus of the Jacobian (\ref{jacobianFermat ; 2.1.2}), which is given by $\Delta=0$, is identical to the discriminant locus of the Fermat-type hypersurface (\ref{hypersurf ; 2.1.2}). 
\par Therefore, the loci given by the following equations in the base 3-fold $\P^1\times\P^1\times\P^1$ describe the discriminant locus of the Fermat-type hypersurface (\ref{hypersurf ; 2.1.2}):
\begin{eqnarray}
\label{discloc ; 2.1.2}
t & = & \alpha_i  \hspace{5mm} (i=1,\cdots, 6) \\ \nonumber
f & = & 0 \\ \nonumber
g & = & 0 \\ \nonumber
h & = & 0.
\end{eqnarray}
Each equation in (\ref{discloc ; 2.1.2}) gives a discriminant component. We use the following notations to denote the discriminant components:
\begin{eqnarray}
A_i  & := & \{t=\alpha_i\}  \hspace{5mm} (i=1,\cdots, 6) \\ \nonumber
B_1 & := & \{f=0\} \\ \nonumber
B_2 & := & \{g=0\} \\ \nonumber
B_3 & := & \{h=0\}.
\end{eqnarray}
We require that
\begin{equation}
B_1\cap B_2\cap B_3=\phi
\end{equation}
to ensure that the Calabi--Yau condition is unbroken. 
\par Component $A_i$, $i=1,\cdots, 6$, is isomorphic to $\P^1\times\P^1$. The bidegree (2,2) curve in $\P^1\times\P^1$ is a curve of genus 1 \footnote{A nonsingular curve of bidegree $(a,b)$ in $\P^1\times \P^1$ is a curve of genus $(a-1)(b-1)$.}, i.e., an elliptic curve $\Sigma_1$, and therefore, component $B_i$, $i=1,2,3$, is isomorphic to $\P^1\times\Sigma_1$. 
\par Next, we determine the intersections of discriminant components; in other words, we find the forms of matter curves that discriminant components contain. When $\alpha_i\ne \alpha_j$, $A_i$ and $A_j$ are parallel. Intersection $A_i\cap B_j$ is a genus-one curve $\Sigma_1$. Two bidegree (2,2) curves in $\P^1\times\P^1$ meet at 8 points \footnote{Two curves of bidegrees $(a,b)$ and $(c,d)$ in $\P^1\times\P^1$ meet at $ad+bc$ points.}, and therefore, $B_i\cap B_j$, $i\ne j$, is a sum of parallel 8 rational curves $\P^1$. We summarize the forms of discriminant components and their intersections in Table \ref{tabFermat ; 2.1.2} below. 

\begingroup
\renewcommand{\arraystretch}{1.5}
\begin{table}[htb]
\begin{center}
  \begin{tabular}{|c|c|} \hline
Component & Topology \\ \hline
$A_i$ & $\P^1\times\P^1$ \\
$B_i$ & $\P^1\times\Sigma_1$ \\ \hline
Intersections &  \\ \hline
$A_i\cap B_j$ & $\Sigma_1$ \\  
$B_i\cap B_j$ & parallel 8 $\P^1$'s \\ \hline 
\end{tabular}
\caption{Discriminant components of Fermat-type hypersurface, and their intersections.}
\label{tabFermat ; 2.1.2}
\end{center}
\end{table}  
\endgroup

\subsubsection{Discriminant Locus and Forms of Discriminant Components of (3,2,2,2) Hypersurfaces in Hesse Form}
\label{sssec 2.1.3}
We determine the discriminant locus and the forms of the discriminant components of (3,2,2,2) hypersurface in Hesse form
\begin{equation}
\label{hypersurf ; 2.1.3}
\begin{split}
(t-\beta_1)(t-\beta_2)aX^3+(t-\beta_3)(t-\beta_4)bY^3+(t-\beta_5)(t-\beta_6)cZ^3 & \\
-3(t-\beta_7)(t-\beta_8)dXYZ & =0.
\end{split}
\end{equation}
We require that all four polynomials $\{a,b,c,d\}$ do not have simultaneous zero, to preserve the Calabi--Yau condition. We also assume that $\beta_7, \beta_8\ne \beta_i$, $i=1,\cdots,6$.
\par We use the following notations
\begin{eqnarray}
A & := & (t-\beta_1)(t-\beta_2)a \\ \nonumber
B & := & (t-\beta_3)(t-\beta_4)b \\ \nonumber
C & := & (t-\beta_5)(t-\beta_6)c \\ \nonumber
D & := & (t-\beta_7)(t-\beta_8)d,
\end{eqnarray}  
and the notation 
\begin{equation}
\begin{split}
F_{Hesse}:= & (t-\beta_1)(t-\beta_2)aX^3+(t-\beta_3)(t-\beta_4)bY^3+(t-\beta_5)(t-\beta_6)cZ^3 \\
 & -3(t-\beta_7)(t-\beta_8)dXYZ.
\end{split}
\end{equation}
Genus-one fiber degenerates exactly when the equations
\begin{equation}
\partial_X F_{Hesse}=\partial_Y F_{Hesse}=\partial_Z F_{Hesse}=0
\end{equation}
have a solution for $[X:Y:Z]\in\P^2$. 
\par From this and by comparing degrees, we obtain the discriminant of the equation (\ref{hypersurf ; 2.1.3}), as follows:
\begin{equation}
\label{discbrev ; 2.1.3}
\Delta=ABC(ABC-D^3)^3
\end{equation}
The discriminant (\ref{discbrev ; 2.1.3}) may be rewritten explicitly as
\begin{equation}
\label{discexp ; 2.1.3}
\Delta=\Pi^6_{i=1}(t-\beta_i)\cdot abc\cdot \big[\Pi^6_{i=1}(t-\beta_i)\cdot abc-(t-\beta_7)^3(t-\beta_8)^3d^3\big]^3.
\end{equation}
We use the notation 
\begin{equation}
e:=\Pi^6_{i=1}(t-\beta_i)\cdot abc-(t-\beta_7)^3(t-\beta_8)^3d^3
\end{equation}
for simplicity. The vanishing of the discriminant $\Delta=0$ describes the discriminant locus. Therefore, the following equations describe the discriminant components:
\begin{eqnarray}
\label{disccomp ; 2.1.3}
t & = & \beta_i \hspace{5mm} (i=1,\cdots,6) \\ \nonumber 
a & = & 0 \\ \nonumber
b & = & 0 \\ \nonumber
c & = & 0 \\ \nonumber
e & = & 0.
\end{eqnarray}
We use the following notations to denote the discriminant components:
\begin{eqnarray}
A_i  & := & \{t=\beta_i\}  \hspace{5mm} (i=1,\cdots, 6) \\ \nonumber
B_1 & := & \{a=0\} \\ \nonumber
B_2 & := & \{b=0\} \\ \nonumber
B_3 & := & \{c=0\} \\ \nonumber
B_4 & := & \{e=0\}.
\end{eqnarray}
\par Component $A_i$ is isomorphic to $\P^1\times\P^1$. The bidegree (2,2) curve in $\P^1\times\P^1$ is a genus-one curve $\Sigma_1$, and therefore, components $B_1, B_2$ and $B_3$ are isomorphic to $\P^1\times\Sigma_1$. $B_4$ is some complicated complex surface. We do not discuss the form of $B_4$. 
\par When $\beta_i\ne\beta_j$, components $A_i$ and $A_j$ are parallel. Intersection $A_i\cap B_j$, $i=1, \cdots,6$, $j=1,\cdots,4$, is isomorphic to $\Sigma_1$. $B_i\cap B_j$, $i,j=1,2,3$, $i\ne j$, is a sum of 8 disjoint rational curves $\P^1$. $B_i\cap B_4$, $i=1,2,3$, is a union of 8 $\P^1$'s and 2 $\Sigma_1$'s. The forms of the discriminant components and their intersections are shown in Table \ref{tabHesse ; 2.1.3} below. 

\begingroup
\renewcommand{\arraystretch}{1.5}
\begin{table}[htb]
\begin{center}
  \begin{tabular}{|c|c|} \hline
Component & Topology \\ \hline
$A_i$ & $\P^1\times\P^1$ \\
$B_i$ ($i=1,2,3$) & $\P^1\times\Sigma_1$ \\ \hline
Intersections &  \\ \hline
$A_i\cap B_j$ ($j=1,\cdots,4$) & $\Sigma_1$ \\ 
$B_i\cap B_j$ ($i,j=1,2,3, i\ne j$) & disjoint 8 $\P^1$'s \\
$B_i\cap B_4$ ($i=1,2,3$) & union of 8 $\P^1$'s and 2 $\Sigma_1$'s \\ \hline
\end{tabular}
\caption{Discriminant components of hypersurfaces in Hesse form, and their intersections. Form of component $B_4$ is omitted.}
\label{tabHesse ; 2.1.3}
\end{center}
\end{table}  
\endgroup

\subsection{Double Covers of $\P^1\times\P^1\times\P^1\times\P^1$ Ramified Along a Multidegree (4,4,4,4) 3-fold}
\subsubsection{Equations for Double Covers of $\P^1\times\P^1\times\P^1\times\P^1$}
Double covers of $\P^1\times\P^1\times\P^1\times\P^1$ ramified along a multidegree (4,4,4,4) 3-fold are Calabi--Yau 4-folds. A fiber of the natural projection onto $\P^1\times\P^1\times\P^1$ is a double cover of $\P^1$ branched along 4 points, which is a genus-one curve. Therefore, projection onto $\P^1\times\P^1\times\P^1$ is a genus-one fibration. Additionally, a fiber of natural projection onto $\P^1\times\P^1$ is a double cover of $\P^1\times\P^1$ branched along a (4,4) curve, which is a genus-one fibered K3 surface; projection onto $\P^1\times\P^1$ gives a K3 fibration. 
\par In this note, we focus on the family of double covers given by the following type of equation:
\begin{equation}
\label{doublecov ; 2.2}
\tau^2=f\cdot a(t)\cdot x^4+g\cdot b(t).
\end{equation}
$x$ is the inhomogeneous coordinate on the first $\P^1$ in the product $\P^1\times\P^1\times\P^1\times\P^1$, and $t$ is the inhomogeneous coordinate on the second $\P^1$. $a$ and $b$ are degree 4 polynomials in the variable $t$. $f$ and $g$ are bidegree (4,4) polynomials on $\P^1\times\P^1$, where the $\P^1$'s in the product $\P^1\times\P^1$ are the last two $\P^1$'s in the product $\P^1\times\P^1\times\P^1\times\P^1$. By splitting the polynomials $a$ and $b$ into linear factors, the equation (\ref{doublecov ; 2.2}) may be rewritten as:
\begin{equation}
\label{factoreddouble ; 2.2}
\tau^2=f\cdot \Pi_{i=1}^4 (t-\alpha_i)\cdot x^4+g\cdot \Pi_{j=5}^8 (t-\alpha_j).
\end{equation}
\par The fiber of the projection onto the product of the third and the fourth $\P^1$'s in $\P^1\times\P^1\times\P^1\times\P^1$ is given by the following equation:
\begin{equation}
\tau^2=\Pi_{i=1}^4 (t-\alpha_i)\cdot x^4+\Pi_{j=5}^8 (t-\alpha_j).
\end{equation}
This is a genus-one fibered K3 surface discussed in \cite{K2}, and it was shown in \cite{K2} that this K3 surface does not admit a global section. Therefore, by a similar argument as that stated in Section \ref{sssec 2.1.1} we conclude that the double covers (\ref{doublecov ; 2.2}) do not have a rational section. 

\subsubsection{Discriminant Locus and Forms of Discriminant Components of Double Covers of $\P^1\times\P^1\times\P^1\times\P^1$}
\label{sssec 2.2.2}
We determine the discriminant locus, and the forms of the discriminant components of double cover (\ref{doublecov ; 2.2}). The Jacobian fibration of double cover (\ref{doublecov ; 2.2}) is given by the following equation \cite{Muk}:
\begin{equation}
\label{jacobian doublecov ; 2.2.2}
\tau^2=\frac{1}{4}x^3-fg\cdot \Pi^8_{i=1}(t-\alpha_i)\cdot x.
\end{equation}
The discriminant of the Jacobian fibration (\ref{jacobian doublecov ; 2.2.2}) is given by 
\begin{equation}
\Delta\sim f^3g^3\cdot \Pi^8_{i=1}(t-\alpha_i)^3. 
\end{equation}  
The condition $\Delta=0$ describes the discriminant locus of the Jacobian (\ref{jacobian doublecov ; 2.2.2}). This is identical to the discriminant locus of double cover (\ref{doublecov ; 2.2}).
\par Therefore, the discriminant locus in the base $\P^1\times\P^1\times\P^1$ is described by the following equations:
\begin{eqnarray}
\label{disccomp ; 2.2.2}
t & = & \alpha_i \hspace{5mm} (i=1,\cdots,8) \\ \nonumber 
f & = & 0 \\ \nonumber
g & = & 0.
\end{eqnarray}
Each equation in (\ref{disccomp ; 2.2.2}) gives a discriminant component. We use the following notations to denote the discriminant components:
\begin{eqnarray}
A_i  & := & \{t=\alpha_i\}  \hspace{5mm} (i=1,\cdots, 8) \\ \nonumber
B_1 & := & \{f=0\} \\ \nonumber
B_2 & := & \{g=0\}.
\end{eqnarray}
\par Discriminant component $A_i$, $i=1,\cdots,8$, is isomorphic to $\P^1\times\P^1$. The bidegree (4,4) curve in $\P^1\times\P^1$ is a genus 9 curve $\Sigma_9$, and therefore, component $B_i$, $i=1,2$, is isomorphic to $\P^1\times\Sigma_9$. 
\par We determine the forms of the intersections of components. When $\alpha_i\ne \alpha_j$, $A_i$ and $A_j$ are parallel. $A_i\cap B_j$, $i=1,\cdots,8$, $j=1,2$, is isomorphic to genus 9 curve $\Sigma_9$. Two bidegree (4,4) curves in $\P^1\times\P^1$ meet at 32 points, and therefore, $B_1\cap B_2$ is the disjoint sum of 32 rational curves $\P^1$. The forms of the discriminant components and their intersections are shown in Table \ref{tab doublecov ; 2.2.2} below.

\begingroup
\renewcommand{\arraystretch}{1.5}
\begin{table}[htb]
\begin{center}
  \begin{tabular}{|c|c|} \hline
Component & Topology \\ \hline
$A_i$ & $\P^1\times\P^1$ \\
$B_i$ & $\P^1\times\Sigma_9$ \\ \hline
Intersections &  \\ \hline
$A_i\cap B_j$ & $\Sigma_9$ \\ 
$B_1\cap B_2$ & disjoint 32 $\P^1$'s \\ \hline
\end{tabular}
\caption{Discriminant components of the double cover of $\P^1\times\P^1\times\P^1\times\P^1$, and their intersections.}
\label{tab doublecov ; 2.2.2}
\end{center}
\end{table}  
\endgroup 

\section{Gauge Symmetries}
\label{sec 3}
We deduce the non-Abelian gauge symmetries that form on the 7-branes in F-theory compactifications on genus-one fibered Calabi--Yau 4-folds lacking a global section, which we constructed in Section \ref{sec 2}. Genus-one fibers of the Fermat-type Calabi--Yau hypersurfaces (\ref{hypersurf 1 ; 2.1.1}) and double covers (\ref{doublecov ; 2.2}) possess complex multiplications of specific orders. These greatly limit the possible monodromies around the singular fibers, and as a result, possible types of singular fibers are also restricted. These strictly constrain the possible non-Abelian gauge groups that can form on the 7-branes. Using this fact, we check the consistency of solutions of non-Abelian gauge groups in Section \ref{ssec 3.4}. Some F-theory models that do not have $U(1)$ gauge symmetry are discussed in Section \ref{ssec 3.5}.

\subsection{Non-Abelian Gauge Groups and Singular Fibers}
When a Calabi--Yau 4-fold has a genus-one fibration, the structures of singular fibers\footnote{See \cite{Kod1, Kod2, Ner, Shiodamodular, Tate, Shioda, Silv, BHPV, SchShio} for discussion of elliptic surfaces, elliptic fibration, and singular fibers. \cite{Cas} discusses elliptic curves and the Jacobian. \cite{Nak, DG, G} discuss elliptic fibrations of 3-folds.} along the codimension one locus in the base are in essence the same as those of singular fibers of elliptic surfaces. Therefore, Kodaira's classification \cite{Kod1, Kod2} applies to singular fibers on discriminant components. According to Kodaira's classification, the types of singular fibers fall into two classes: i) six types $II$, $III$, $IV$, $II^*$, $III^*$, and $IV^*$; and ii) two infinite series $I_n$ ($n\ge 1$) and $I^*_m$ ($m\ge 0$).
\par
\begin{figure}
\begin{center}
\includegraphics[height=10cm, bb=0 0 960 720]{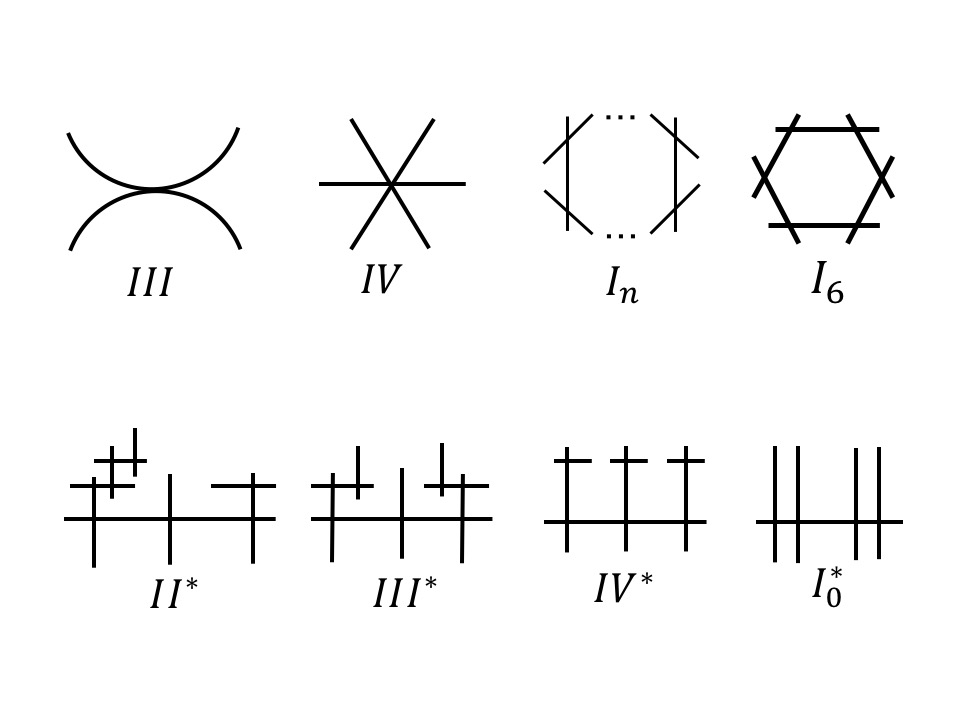}
\caption{\label{figurefiber ; 3.1}Singular Fibers}
\end{center}
\end{figure}

 Fibers of type $I_1$ and $II$ are rational curves $\P^1$ with one singularity ($II$ is a rational curve with a cusp, and $I_1$ is a rational curve with a node); fibers of the other types are unions of smooth $\P^1$'s intersecting in specific ways. Type $III$ fiber is a union of two rational curves tangential to each other at one point, and type $IV$ fiber is a union of three rational curves meeting at one point. For each fiber type $I_n$, $n$ rational curves intersect to form an $n$-gon. Figure \ref{figurefiber ; 3.1} shows images of the singular fibers. Each line in the image represents a rational curve $\P^1$. Two rational curve components in a singular fiber intersect only when two lines in an image intersect. 

\par Non-Abelian gauge group that forms on the 7-branes is determined by the singular fiber type over a discriminant component. The correspondence between the non-Abelian gauge symmetries on the 7-branes and the fiber types is discussed in \cite{MV2, BIKMSV}. The correspondences of the types of singular fibers and the singularity types are presented in Table \ref{tabgauge ; 3.1} below. 

\begingroup
\renewcommand{\arraystretch}{1.5}
\begin{table}[htb]
\begin{center}
  \begin{tabular}{|c|c|} \hline
fiber type & singularity type \\ \hline
$I_n$ ($n\ge 2$) & $A_{n-1}$ \\
$I^*_m$ ($m\ge 0$) & $D_{4+m}$ \\
$III$ & $A_1$ \\
$IV$ & $A_2$ \\
$II^*$ & $E_8$ \\
$III^*$ & $E_7$ \\
$IV^*$ & $E_6$ \\ \hline   
\end{tabular}
\caption{Correspondence between singular fiber types and singularity types.}
\label{tabgauge ; 3.1}
\end{center}
\end{table}  
\endgroup 

\subsection{Non-Abelian Gauge Groups in F-theory on (3,2,2,2) Hypersurfaces}
\label{ssec 3.2}
We deduce non-Abelian gauge symmetries in F-theory compactification on (3,2,2,2) hypersurfaces. 
\subsubsection{Fermat-Type (3,2,2,2) Hypersurfaces}
\label{sssec 3.2.1}
We deduce the non-Abelian gauge symmetries that form on the 7-branes in F-theory compactifications on the Fermat-type (3,2,2,2) Calabi--Yau hypersurfaces
\begin{equation}
(t-\alpha_1)(t-\alpha_2)fX^3+(t-\alpha_3)(t-\alpha_4)gY^3+(t-\alpha_5)(t-\alpha_6)hZ^3=0.
\label{hypersurf ; 3.2.1}
\end{equation}
As stated in Section \ref{sssec 2.1.2}, the Jacobian fibration of Fermat-type hypersurface (\ref{hypersurf ; 3.2.1}) is given by the following equation:
\begin{equation}
\label{jacobianFermat ; 3.2.1}
X^3+Y^3+\Pi_{i=1}^6 (t-\alpha_i)\cdot fgh\cdot Z^3=0.
\end{equation}  
The Jacobian fibration (\ref{jacobianFermat ; 3.2.1}) transforms into the following Weierstrass form:  
\begin{equation}
\label{localjacobian ; 3.2.1}
y^2=x^3-2^4\cdot 3^3\cdot \Pi_{i=1}^6 (t-\alpha_i)^2\cdot f^2g^2h^2, 
\end{equation}
and the discriminant of the Jacobian fibration (\ref{jacobianFermat ; 3.2.1}) is given by the following equation:
\begin{equation}
\Delta \sim \Pi_{i=1}^6 (t-\alpha_i)^4\cdot f^4g^4h^4.
\label{discriminant ; 3.2.1}
\end{equation} 
\par Fermat-type (3,2,2,2) hypersurface (\ref{hypersurf ; 3.2.1}) and the Jacobian fibration (\ref{jacobianFermat ; 3.2.1}) have the identical singular fiber types over the same discriminant loci, thus the result of singular fibers for the Jacobian fibration (\ref{jacobianFermat ; 3.2.1}) gives identical singular fibers of Fermat-type hypersurface (\ref{hypersurf ; 3.2.1}). 
\par We determine the types of singular fibers of the Jacobian fibration (\ref{jacobianFermat ; 3.2.1}) from the Weierstrass form (\ref{localjacobian ; 3.2.1}) and the discriminant (\ref{discriminant ; 3.2.1}). We show the correspondence of the singular fiber types and the vanishing orders of the coefficients of the Weierstrass form in Table \ref{tabcoeff ; 3.2.1}.
\begingroup
\renewcommand{\arraystretch}{1.5}
\begin{table}[htb]
\begin{center}
  \begin{tabular}{|c|c|c|c|} \hline
Fiber type & Ord($a_4$) & Ord($a_6$) & Ord($\Delta$) \\ \hline
$I_0$ & $\ge 0$ & $\ge 0$ & 0 \\ \hline
$I_n$  ($n\ge 1$) & 0 & 0 & $n$ \\ \hline
$II$ & $\ge 1$ & 1 & 2 \\ \hline
$III$ & 1 & $\ge 2$ & 3 \\ \hline
$IV$ & $\ge 2$ & 2 & 4 \\ \hline
$I_0^*$ & 2 & $\ge 3$ & 6 \\ \cline{2-4}
 & $\ge 2$ & 3 & 6 \\ \hline
$I_m^*$  ($m \ge 1$) & 2 & 3 & $6+m$ \\ \hline
$IV^*$ & $\ge 3$ & 4 & 8 \\ \hline
$III^*$ & 3 & $\ge 5$ & 9 \\ \hline
$II^*$ & $\ge 4$ & 5 & 10 \\ \hline   
\end{tabular}
\caption{\label{tabcoeff ; 3.2.1}Correspondence of the types of singular fibers and the vanishing orders of coefficients $a_4, a_6$, and the discriminant $\Delta$, of the Weierstrass form $y^2=x^3+a_4x+a_6$.}
\end{center}
\end{table}  
\endgroup 
We find that, when $\alpha_i$ ($i=1,\cdots,6$) are mutually distinct, the singular fiber on component $A_i$ is of type $IV$. The polynomial 
\begin{equation}
y^2+2^4\cdot 3^3\cdot \Pi_{i=1}^6 (t-\alpha_i)^2\cdot f^2g^2h^2
\label{polynomial ; 3.2.1}
\end{equation}
splits into linear factors as 
\begin{equation}
(y+2^2\cdot 3\sqrt{3}i\cdot \Pi_{i=1}^6 (t-\alpha_i)\cdot fgh)(y-2^2\cdot 3\sqrt{3}i\cdot \Pi_{i=1}^6 (t-\alpha_i)\cdot fgh).
\label{split ; 3.2.1}
\end{equation}
Thus, we find that type $IV$ fiber on component $A_i$ is of split type \cite{BIKMSV}. Therefore, $SU(3)$ gauge symmetry arises on 7-branes wrapped on component $A_i$. When the multiplicity of $\alpha_i$ is 2, (i.e. when there is one $j\ne i$ such that $\alpha_i=\alpha_j$), 7-branes wrapped on components $A_i$ and $A_j$ coincide, and the fiber type is enhanced to $IV^*$. Since polynomial (\ref{polynomial ; 3.2.1}) splits into linear factors as (\ref{split ; 3.2.1}), we find that type $IV^*$ fiber on component $A_i$ is split. The corresponding gauge group on 7-branes is enhanced to $E_6$. To preserve Calabi--Yau condition, the multiplicity cannot be greater than 2. Type of singular fibers on component $B_i$ is $IV$, and we see that they are of split type from factorization (\ref{split ; 3.2.1}); $SU(3)$ gauge symmetry arises on 7-branes wrapped on component $B_i$. The results are summarized in Table \ref{tabFermat gauge ; 3.2.1} below. 

\begingroup
\renewcommand{\arraystretch}{1.5}
\begin{table}[htb]
\begin{center}
  \begin{tabular}{|c|c|c|} \hline
Component & Fiber type & non-Abel. Gauge Group \\ \hline
$A_i$ & $IV$ & $SU(3)$ \\ \cline{2-3}
      & $IV^*$ & $E_6$ \\ \hline
$B_i$ & $IV$ & $SU(3)$ \\ \hline   
\end{tabular}
\caption{Types of singular fibers and corresponding non-Abelian gauge groups on discriminant components of Fermat-type hypersurface.}
\label{tabFermat gauge ; 3.2.1}
\end{center}
\end{table}  
\endgroup        

\subsubsection{(3,2,2,2) Hypersurfaces in Hesse Form}
We determine the types of singular fibers of (3,2,2,2) hypersurfaces in Hesse form 
\begin{equation}
\begin{split}
(t-\beta_1)(t-\beta_2)aX^3+(t-\beta_3)(t-\beta_4)bY^3+(t-\beta_5)(t-\beta_6)cZ^3 & \\
-3(t-\beta_7)(t-\beta_8)dXYZ & =0
\end{split}
\label{hypersurf 2 ; 3.2.2}
\end{equation}
by computing the singular fibers of the Jacobian fibration. As we saw in Section \ref{sssec 2.1.3}, the equation for (3,2,2,2) hypersurface in Hesse form (\ref{hypersurf 2 ; 3.2.2}) has the following discriminant:
\begin{equation}
\label{discexp ; 3.2.2}
\Delta=\Pi^6_{i=1}(t-\beta_i)\cdot abc\cdot e^3.
\end{equation}
In (\ref{discexp ; 3.2.2}), we have used the notation 
\begin{equation}
e=\Pi^6_{i=1}(t-\beta_i)\cdot abc-(t-\beta_7)^3(t-\beta_8)^3d^3
\end{equation}
for simplicity. 
\par The Jacobian fibration of (3,2,2,2) hypersurface in Hesse form (\ref{hypersurf 2 ; 3.2.2}) is given as:  
\begin{equation}
\label{jacobianhypersurf2 long ; 3.2.2}
X^3+Y^3+\Pi^6_{i=1}(t-\beta_i)\cdot abc\cdot Z^3-3(t-\beta_7)(t-\beta_8)dXYZ=0.
\end{equation}
The discriminant of the Jacobian fibration (\ref{jacobianhypersurf2 long ; 3.2.2}) of (3,2,2,2) hypersurface in Hesse form (\ref{hypersurf 2 ; 3.2.2}) is also given by (\ref{discexp ; 3.2.2}).
\par As in Section \ref{sssec 2.1.3}, we use the following notations:
\begin{eqnarray}
\label{notation ; 3.2.2}
A & := & (t-\beta_1)(t-\beta_2)a \\ \nonumber
B & := & (t-\beta_3)(t-\beta_4)b \\ \nonumber
C & := & (t-\beta_5)(t-\beta_6)c \\ \nonumber
D & := & (t-\beta_7)(t-\beta_8)d,
\end{eqnarray}  
Using the notations (\ref{notation ; 3.2.2}), (3,2,2,2) hypersurface in Hesse form (\ref{hypersurf 2 ; 3.2.2}) may be rewritten as:
\begin{equation}
AX^3+BY^3+CZ^3-3D\cdot XYZ=0.
\label{short hypersurf 2 ; 3.2.2}
\end{equation}
The Jacobian fibration (\ref{jacobianhypersurf2 long ; 3.2.2}) of (3,2,2,2) hypersurface in Hesse form (\ref{hypersurf 2 ; 3.2.2}) may be rewritten as: 
\begin{equation}
X^3+Y^3+ABCZ^3-3D\cdot XYZ=0.
\label{jacobian hypersurf 2 ; 3.2.2}
\end{equation}
Using the notations (\ref{notation ; 3.2.2}), both the discriminants of (3,2,2,2) hypersurface in Hesse form (\ref{short hypersurf 2 ; 3.2.2}) and the Jacobian fibration (\ref{jacobian hypersurf 2 ; 3.2.2}) are given as follows:
\begin{equation}
\Delta=ABC(ABC-D^3)^3.
\end{equation}
\par Jacobian fibration (\ref{jacobian hypersurf 2 ; 3.2.2}) transforms into the general Weierstrass form as
\begin{equation}
y^2-3Dxy+(ABC-D^3)y=x^3.
\label{general Weierstrass ; 3.2.2}
\end{equation}
We complete the square in $y$ as $\til{y}=y+\frac{1}{2}(-3Dx+ABC-D^3)$, and complete the cube in $x$ as $\til{x}=x+\frac{3}{4}D^2$ to obtain the following Weierstrass form:
\begin{equation}
\til{y}^2=\til{x}^3-(\frac{3}{2}ABCD+\frac{3}{16}D^4)\til{x}+(\frac{1}{4}(ABC)^2+\frac{5}{8}ABCD^3-\frac{1}{32}D^6).
\label{Weierstrass hypersurf ; 3.2.2}
\end{equation}
\par We deduce from Weierstrass form (\ref{Weierstrass hypersurf ; 3.2.2}) that type of singular fibers over each discriminant component is $I_n$ for some $n\ge 1$. Therefore, the types of singular fibers can be determined by studying the orders of the zeros of the discriminant (\ref{discexp ; 3.2.2}). In the general Weierstrass form (\ref{general Weierstrass ; 3.2.2}), polynomial
\begin{equation}
y^2-3Dxy
\end{equation}
can be factored as 
\begin{equation}
y(y-3Dx).
\end{equation}
Thus, we conclude that singular fibers on component $B_4$ are of split type. 
\par Under the translation in $x$ and $y$ that replaces $x$ with $x-D^2$, and $y$ with $y-D^3$, the general Weierstrass form (\ref{general Weierstrass ; 3.2.2}) transforms into another general Weierstrass form:
\begin{equation}
y^2-3Dxy+ABCy=x^3-3D^2x^2+ABCD^3.
\end{equation}
Polynomial
\begin{equation}
y^2-3Dxy+3D^2x^2
\end{equation}
splits into linear factors as 
\begin{equation}
(y-\frac{1}{2}(3-i\sqrt{3})Dx) (y-\frac{1}{2}(3+i\sqrt{3})Dx).
\end{equation}
Therefore, we deduce that singular fibers on components $A_i$, $i=1,2,\cdots, 6$, are split. Non-Abelian gauge groups that form on the 7-branes wrapped on components $A_i$, $i=1,2, \cdots, 6$, and component $B_4$, are of the form $SU(N)$.
\par When $\beta_i$'s are mutually distinct, the fiber type on component $A_i$ is $I_1$, and non-Abelian gauge symmetry does not form on the 7-brane wrapped on $A_i$. As the multiplicity of $\beta_i$ increases, more 7-branes become coincident, and the non-Abelian gauge group becomes further enhanced. The maximum enhancement occurs when all $\beta_i$, $i=1,\cdots,6$, are equal, and all six 7-branes wrapped on $A_i$ coincide. The fiber type on component $A_1$ for this case is $I_6$, and $SU(6)$ gauge symmetry arises on the 7-branes wrapped on $A_1$. In Section \ref{sec 5}, we compute the potential matter spectra for this most enhanced situation.
\par Singular fibers on component $B_i$, $i=1,2,3$, have type $I_1$; a non-Abelian gauge group does not form on the 7-brane wrapped on component $B_i$, $i=1,2,3$. Singular fibers on component $B_4$ have type $I_3$, and $SU(3)$ gauge group arises on 7-branes wrapped on $B_4$. Results are summarized in Table \ref{tabHesse gauge ; 3.2.2}.

\begingroup
\renewcommand{\arraystretch}{1.5}
\begin{table}[htb]
\begin{center}
  \begin{tabular}{|c|c|c|} \hline
Component & Fiber type & non-Abel. Gauge Group \\ \hline
      & $I_1$ & None. \\ \cline{2-3}
      & $I_2$ & $SU(2)$ \\ \cline{2-3}
$A_i$ & $I_3$ & $SU(3)$ \\ \cline{2-3}
      & $I_4$ & $SU(4)$ \\ \cline{2-3}
      & $I_5$ & $SU(5)$ \\ \cline{2-3}
      & $I_6$ & $SU(6)$ \\ \hline
$B_{1,2,3}$ & $I_1$ & None. \\ 
$B_4$ & $I_3$ & $SU(3)$ \\ \hline   
\end{tabular}
\caption{Types of singular fibers and corresponding non-Abelian gauge groups on discriminant components of hypersurface in Hesse form.}
\label{tabHesse gauge ; 3.2.2}
\end{center}
\end{table}  
\endgroup          

\subsection{Non-Abelian Gauge Groups in F-theory on Double Covers of $\P^1\times\P^1\times\P^1\times\P^1$}
\label{ssec 3.3}
We deduce the non-Abelian gauge groups in F-theory compactifications on double covers
\begin{equation}
\label{factoreddouble ; 3.3}
\tau^2=f\cdot \Pi_{i=1}^4 (t-\alpha_i)\cdot x^4+g\cdot \Pi_{j=5}^8 (t-\alpha_j).
\end{equation}
\par As stated in Section \ref{sssec 2.2.2}, the Jacobian fibration of double cover (\ref{factoreddouble ; 3.3}) is given by the following equation:
\begin{equation}
\label{jacobian doublecov ; 3.3}
\tau^2=\frac{1}{4}x^3-fg\cdot \Pi^8_{i=1}(t-\alpha_i)\cdot x.
\end{equation}
The discriminant of the Jacobian fibration (\ref{jacobian doublecov ; 3.3}) is given by 
\begin{equation}
\Delta\sim f^3g^3\cdot \Pi^8_{i=1}(t-\alpha_i)^3. 
\end{equation}  
\par We determine the types of singular fibers of double cover (\ref{factoreddouble ; 3.3}) by computing the types of singular fibers of the Jacobian fibration (\ref{jacobian doublecov ; 3.3}). When $\alpha_i$'s are mutually distinct, the singular fiber on component $A_i$ has type $III$; the $SU(2)$ gauge group arises on the 7-branes wrapped on component $A_i$ for this case. When the multiplicity of $\alpha_i$ is 2, say there is $j\ne i$ such that $\alpha_i=\alpha_j$, then the 7-branes wrapped on components $A_i$ and $A_j$ become coincident, and singular fiber on component $A_i$ has type $I^*_0$. The polynomial 
\begin{equation}
x^3-fg\cdot x
\end{equation}
splits into the quadratic factor and the linear factor as 
\begin{equation}
x(x^2-fg)
\end{equation}
for generic polynomials $f,g$. Therefore, we conclude that $I^*_0$ fiber on component $A_i$ is semi-split; the non-Abelian gauge symmetry on the 7-branes wrapped on component $A_i$ becomes enhanced to $SO(7)$. When the multiplicity of $\alpha_i$ is 3, the singular fiber on component $A_i$ has type $III^*$, and the gauge symmetry on component $A_i$ is further enhanced to $E_7$. To preserve the Calabi--Yau condition, no further enhancement is possible. The singular fibers on component $B_i$ is of type $III$; the $SU(2)$ gauge group arises on 7-branes wrapped on component $B_i$. The results are displayed in Table \ref{tab doublecov gauge ; 3.3} below.

\begingroup
\renewcommand{\arraystretch}{1.5}
\begin{table}[htb]
\begin{center}
  \begin{tabular}{|c|c|c|} \hline
Component & Fiber type & non-Abel. Gauge Group \\ \hline
      & $III$ & $SU(2)$ \\ \cline{2-3}
$A_i$ & $I^*_0$ & $SO(7)$ \\ \cline{2-3}      
      & $III^*$ & $E_7$ \\ \hline
$B_i$ & $III$ & $SU(2)$ \\ \hline   
\end{tabular}
\caption{Types of singular fibers and corresponding non-Abelian gauge groups on discriminant components of double cover of $\P^1\times\P^1\times\P^1\times\P^1$ (\ref{factoreddouble ; 3.3}).}
\label{tab doublecov gauge ; 3.3}
\end{center}
\end{table}  
\endgroup   

\subsection{Consistency Check by Monodromy}
\label{ssec 3.4}
We consider monodromies around singular fibers to perform a consistency check of solutions of non-Abelian gauge groups, which we obtained in Sections \ref{sssec 3.2.1}, \ref{ssec 3.3}. Genus-one fibers of Fermat-type (3,2,2,2) hypersurfaces (\ref{hypersurf ; 3.2.1}) and double covers (\ref{factoreddouble ; 3.3}) possess particular symmetries, and as a result, these symmetries strictly constrain monodromies around singular fibers. We confirm that the non-Abelian gauge symmetries obtained by us in agreement with these restrictions. 

\subsubsection{Monodromy and J-invariant}
\label{sssec 3.4.1}
Genus-one fibers of Fermat-type (3,2,2,2) hypersurfaces (\ref{hypersurf ; 3.2.1}) and double covers (\ref{factoreddouble ; 3.3}) have constant j-invariants; they are constant over the base 3-fold $\P^1\times\P^1\times\P^1$. 
\par Concretely, generic genus-one fiber of the Fermat-type (3,2,2,2) hypersurface is the Fermat curve\footnote{The Fermat curve possesses complex multiplication of order 3.}, whose j-invariant is known to be 0. Therefore, the j-invariant of singular fibers is forced to be 0. 
\par Smooth genus-one fiber of double cover (\ref{factoreddouble ; 3.3}) is invariant under the map:
\begin{equation}
x\rightarrow e^{\rm 2\pi i/4}x,
\end{equation}
whose order is 4. This is a complex multiplication of order 4, and therefore, the generic genus-one fiber has the j-invariant 1728. This forces the j-invariant of singular fibers to be 1728. 
\par Each fiber type has a specific monodromy and j-invariant. We display the monodromy and their orders in $SL_2(\Z)$, and the j-invariant, for each fiber type in Table \ref{tabmonodromy ; 3.4} below. ``Finite'' in the table means that the j-invariant of fiber type $I^*_0$ can take any finite value in $\C$. Results in Table \ref{tabmonodromy ; 3.4} were derived in \cite{Kod1, Kod2}\footnote{Euler numbers of fiber types were obtained in \cite{Kod2}, and they have an interpretation as the number of 7-branes wrapped on.}.

\begingroup
\renewcommand{\arraystretch}{1.1}
\begin{table}[htb]
  \begin{tabular}{|c|c|r|c|c|} \hline
Fiber Type & j-invariant & Monodromy  & order of Monodromy & \# of 7-branes (Euler number) \\ \hline
$I^*_0$ & finite & $-\begin{pmatrix}
1 & 0 \\
0 & 1 \\
\end{pmatrix}$ & 2 & 6\\ \hline
$I_b$ & $\infty$ & $\begin{pmatrix}
1 & b \\
0 & 1 \\
\end{pmatrix}$ & infinite & $b$\\
$I^*_b$ & $\infty$ & $-\begin{pmatrix}
1 & b \\
0 & 1 \\
\end{pmatrix}$ & infinite & $b+$6\\ \hline
$II$ & 0 & $\begin{pmatrix}
1 & 1 \\
-1 & 0 \\
\end{pmatrix}$ & 6 & 2\\
$II^*$ & 0 & $\begin{pmatrix}
0 & -1 \\
1 & 1 \\
\end{pmatrix}$ & 6 & 10\\ \hline
$III$ & 1728 & $\begin{pmatrix}
0 & 1 \\
-1 & 0 \\
\end{pmatrix}$ & 4 & 3\\
$III^*$ & 1728 & $\begin{pmatrix}
0 & -1 \\
1 & 0 \\
\end{pmatrix}$ & 4 & 9\\ \hline
$IV$ & 0 & $\begin{pmatrix}
0 & 1 \\
-1 & -1 \\
\end{pmatrix}$ & 3 & 4\\
$IV^*$ & 0 & $\begin{pmatrix}
-1 & -1 \\
1 & 0 \\
\end{pmatrix}$ & 3 & 8\\ \hline
\end{tabular}
\caption{Fiber types, their j-invariants, monodromies, and the associated numbers of 7-branes.}
\label{tabmonodromy ; 3.4}
\end{table}
\endgroup 

\subsubsection{Fermat-Type (3,2,2,2) Hypersurfaces}
As we saw in Section \ref{sssec 3.4.1}, singular fibers of the Fermat-type (3,2,2,2) hypersurface have j-invariant 0. As can be seen in Table \ref{tabmonodromy ; 3.4}, the fiber types with j-invariant 0 are only $II$, $IV$, $I^*_0$, $IV^*$, and $II^*$. (j-invariant of type $I^*_0$ fiber can take the value 0.) Fiber types on discriminant components that we obtained in Section \ref{sssec 3.2.1} are $IV, IV^*$, which is in agreement with constraint imposed by the j-invariant. Monodromies of order 3 characterize non-Abelian gauge symmetries arising on 7-branes in F-theory compactifications on Fermat-type (3,2,2,2) hypersurfaces.

\subsubsection{Double Covers of $\P^1\times\P^1\times\P^1\times\P^1$}
As we saw in Section \ref{sssec 3.4.1}, singular fibers of double cover (\ref{factoreddouble ; 3.3}) have j-invariant 1728. According to Table \ref{tabmonodromy ; 3.4}, fiber types with j-invariant 1728 are only $III$, $I^*_0$, and $III^*$. This agrees with the fiber types that we obtained in Section \ref{ssec 3.3} on discriminant components of double covers. Monodromies of order 2 and 4 characterize non-Abelian gauge symmetries on 7-branes in F-theory compactification on double covers (\ref{factoreddouble ; 3.3}).

\subsection{F-theory Models without $U(1)$ Symmetry}
\label{ssec 3.5}
We specify the Mordell--Weil groups of the Jacobian fibrations of special genus-one fibered Calabi--Yau 4-folds. We find that the Mordell--Weil groups of Jacobian fibrations of the special genus-one fibered Calabi--Yau 4-folds that we consider here have the rank 0, therefore, we deduce that F-theory compactifications on these special genus-one fibered Calabi--Yau 4-folds do not have a $U(1)$ gauge symmetry. 
\subsubsection{Special Fermat-Type (3,2,2,2) Hypersurface}
\label{sssec 3.5.1}
\par We particularly consider the following special Fermat-type (3,2,2,2) hypersurface:
\begin{equation}
(t-\alpha_1)^2fX^3+(t-\alpha_2)^2gY^3+(t-\alpha_3)^2hZ^3=0.
\label{hypersurf ; 3.5}
\end{equation}
The Jacobian fibration of this special Fermat-type hypersurface (\ref{hypersurf ; 3.5}) is given by the following equation:
\begin{equation}
X^3+Y^3+(t-\alpha_1)^2(t-\alpha_2)^2(t-\alpha_3)^2\cdot fgh\cdot Z^3=0.
\label{jacobianhypersurf ; 3.5}
\end{equation}
The projection onto the last two $\P^1$'s in $\P^2\times\P^1\times\P^1\times\P^1$ gives a K3 fibration, and picking a point in the base surface $\P^1\times\P^1$ gives a specialization to this K3 fiber. The K3 fiber of the Jacobian fibration (\ref{jacobianhypersurf ; 3.5}) is given by the following equation:
\begin{equation}
X^3+Y^3+(t-\alpha_1)^2(t-\alpha_2)^2(t-\alpha_3)^2Z^3=0.
\label{fibhypersurf ; 3.5}
\end{equation}
This is the Jacobian fibration of the Fermat-type K3 hypersurface (\ref{K3 1 ; 2.1.1}), which is discussed in \cite{K}, with reducible fiber type $E^3_6$. According to Table 2 in \cite{SZ}, extremal K3 surface \footnote{Extremal K3 surface is an elliptic K3 surface with a section having the Picard number 20, with the Mordell--Weil rank 0.} with reducible fiber type $E^3_6$ is uniquely determined, and its transcendental lattice has the intersection matrix $\begin{pmatrix}
2 & 1 \\
1 & 2 \\
\end{pmatrix}.$
The Mordell--Weil group of this extremal K3 surface is determined in \cite{Nish, SZ} to be $\Z_3$. 
\par By considering the specialization of the Jacobian fibration (\ref{jacobianhypersurf ; 3.5}) to its K3 fiber (\ref{fibhypersurf ; 3.5}), we find that the Mordell--Weil group of the Jacobian (\ref{jacobianhypersurf ; 3.5}) is isomorphic to that of its K3 fiber (\ref{fibhypersurf ; 3.5}), which is $\Z_3$. This shows that the Mordell--Weil group of the Jacobian fibration (\ref{jacobianhypersurf ; 3.5}) is isomorphic to $\Z_3$. Thus, we conclude that the global structure of the non-Abelian gauge group in F-theory compactified on the special Fermat-type (3,2,2,2) hypersurface (\ref{hypersurf ; 3.5}) is given by the following:
\begin{equation}
E^3_6 \times SU(3)^3 / \Z_3.
\end{equation}
\par In particular, the Mordell--Weil group of the Jacobian fibration (\ref{jacobianhypersurf ; 3.5}) has rank 0, therefore, F-theory compactified on the Fermat-type (3,2,2,2) hypersurface (\ref{hypersurf ; 3.5}) does not have a $U(1)$ gauge symmetry. 

\subsubsection{Special Double Cover of $\P^1\times\P^1\times\P^1\times\P^1$}
\par Next, we consider the double cover of $\P^1\times\P^1\times\P^1\times\P^1$ given by the following equation:
\begin{equation}
\label{doublecov ; 3.5}
\tau^2=f\cdot (t-\alpha_1)^3(t-\alpha_2)\cdot x^4+g\cdot (t-\alpha_2)(t-\alpha_3)^3.
\end{equation}
The Jacobian fibration of this double cover is given by:
\begin{equation}
\label{jacobian doublecov ; 3.5}
\tau^2=\frac{1}{4}x^3-fg\cdot (t-\alpha_1)^3(t-\alpha_2)^2(t-\alpha_3)^3\cdot x.
\end{equation}
The K3 fiber of the Jacobian fibration (\ref{jacobian doublecov ; 3.5}) is given by the equation:
\begin{equation}
\label{fibjacobian doublecov ; 3.5}
\tau^2=\frac{1}{4}x^3-(t-\alpha_1)^3(t-\alpha_2)^2(t-\alpha_3)^3\cdot x.
\end{equation}
K3 surface (\ref{fibjacobian doublecov ; 3.5}) is extremal K3 with the reducible fiber type $E^2_7D_4$. As discussed in \cite{K2}, the Mordell--Weil group of this extremal K3 surface (\ref{fibjacobian doublecov ; 3.5}) is $\Z_2$ \cite{Nish, SZ}. 
\par As per reasoning similar to the argument in Section \ref{sssec 3.5.1}, we consider the specialization of the Jacobian fibration of the double cover (\ref{jacobian doublecov ; 3.5}) to its K3 fiber (\ref{fibjacobian doublecov ; 3.5}) and find that the Mordell--Weil group of the Jacobian (\ref{jacobian doublecov ; 3.5}) is isomorphic to that of its K3 fiber (\ref{fibjacobian doublecov ; 3.5}). Therefore, we conclude that the Mordell--Weil group of the Jacobian fibration (\ref{jacobian doublecov ; 3.5}) is isomorphic to $\Z_2$. Thus, we deduce that the global structure of the non-Abelian gauge group in F-theory compactification on the special double cover (\ref{doublecov ; 3.5}) is given by the following:
\begin{equation}
E^2_7\times SO(7) \times SU(2)^2 / \Z_2.
\end{equation}
\par The Mordell--Weil group of the Jacobian (\ref{jacobian doublecov ; 3.5}) has rank 0, therefore, it follows that F-theory compactification on the double cover (\ref{doublecov ; 3.5}) does not have a $U(1)$ gauge symmetry. 

\section{Discussion of Consistent Four-Form Flux and Euler Characteristics of Calabi--Yau 4-folds}
\label{sec 4}
\subsection{Review of Conditions on Four-Form Flux}
We briefly review physical conditions imposed on four-form flux $G_4$ of genus-one fibered Calabi--Yau 4-fold $Y$. The quantization condition \cite{W} imposed on four-form flux is given by the following equation:
\begin{equation}
G_4+\frac{1}{2}c_2(Y)\in H^4(Y,\Z).
\end{equation}
In particular, when the second Chern class $c_2(Y)$ is even, the term $\frac{1}{2}c_2(Y)$ is irrelevant. To preserve supersymmetry in 4d theory, the following conditions need to be imposed \cite{BB} on four-form flux:
\begin{equation}
\label{2,2cond ; 4.1}
G_4\in H^{2,2}(Y)
\end{equation}
\begin{equation}
\label{prim ; 4.1}
G_4\wedge J=0.
\end{equation}
$J$ in the condition (\ref{prim ; 4.1}) represents a K{\" a}hler form. 
\par Furthermore, to ensure that the 4d effective theory has Lorentz symmetry, four-form flux is required to have one leg in the fiber \cite{DRS}. When genus-one fibration admits a global section, this condition is given by the following equations:
\begin{equation}
G_4\cdot \til{p}^{-1}(C)\cdot \til{p}^{-1}(C')=0
\label{intersec ; 4.1}
\end{equation}
\begin{equation}
\label{fibersec ; 4.1}
G_4\cdot S_0 \cdot \til{p}^{-1}(C)=0
\end{equation}
for any $C,C'\in H^{1,1}(B_3)$. $B_3$ denotes base 3-fold. In the equations (\ref{intersec ; 4.1}) and (\ref{fibersec ; 4.1}), $\til{p}$ denotes the projection from elliptically fibered Calabi--Yau 4-fold $Y$ onto base 3-fold $B_3$. In the equation (\ref{fibersec ; 4.1}), $S_0$ denotes a rational zero section. 
\par Generalization of the conditions (\ref{intersec ; 4.1}) and (\ref{fibersec ; 4.1}) to genus-one fibration without a section was proposed in \cite{LMTW}; the generalized equations are as follows:
\begin{equation}
\label{genusonefib ; 4.1}
G_4\cdot p^{-1}(C)\cdot p^{-1}(C')=0
\end{equation}
\begin{equation}
G_4\cdot \hat{N} \cdot p^{-1}(C)=0
\label{genusonemult ; 4.1}
\end{equation}
for any $C,C'\in H^{1,1}(B_3)$. In the equations (\ref{genusonefib ; 4.1}) and (\ref{genusonemult ; 4.1}), $p$ denotes the projection from genus-one fibered Calabi--Yau 4-fold $Y$ onto base 3-fold $B_3$. $\hat{N}$ is some appropriate sum of an $n$-section $N$ that Calabi--Yau genus-one fibration $Y$ possesses and exceptional divisors. 
\par The condition to cancel the tadpole, including 3-branes, is given as follows \cite{VW, SVW}:
\begin{equation}
\frac{\chi(Y)}{24}=\frac{1}{2}G_4\cdot G_4+N_3.
\end{equation}
$N_3$ denotes the number of 3-branes minus anti 3-branes, and the stability of compactification requires $N_3\ge 0$. 

\subsection{Intrinsic Algebraic 2-cycles as Candidates for Four-Form Fluxes}
\label{ssec 4.2}
We use algebraic 2-cycles as candidates for four-form fluxes. With this choice, the condition (\ref{2,2cond ; 4.1}) is satisfied. 
\par We refer to algebraic 2-cycles of (3,2,2,2) hypersurfaces as the {\it intrinsic} algebraic 2-cycles of (3,2,2,2) hypersurfaces in this study, when they do not belong to the algebraic 2-cycles obtained as the restrictions of algebraic cycles in the ambient space $\P^2\times\P^1\times\P^1\times\P^1$. Similarly, we refer to the algebraic 2-cycles of double covers of $\P^1\times\P^1\times\P^1\times\P^1$ as the intrinsic algebraic 2-cycles, when they do not belong to the algebraic 2-cycles obtained as the pullbacks of algebraic cycles of the product $\P^1\times\P^1\times\P^1\times\P^1$.
\par We show that the nonintrinsic algebraic 2-cycles of a (3,2,2,2) hypersurface, namely the algebraic 2-cycles obtained as the restrictions of algebraic cycles in $\P^2\times\P^1\times\P^1\times\P^1$, do not yield consistent four-form fluxes. This can be shown as follows: an algebraic 2-cycle obtained as the restriction of an algebraic cycle in the product $\P^2\times\P^1\times\P^1\times\P^1$ is given as follows:
\begin{equation}
\label{2-cycle in 4.2}
(\alpha_1\, x^2+\alpha_2 \, xy+\alpha_3\, xz+\alpha_4\, xw+\alpha_5\, yz+\alpha_6\, yw+\alpha_7\, zw)|_Y.
\end{equation}
We used $|_Y$ to denote the restriction to Calabi--Yau (3,2,2,2) hypersurface $Y$. $\alpha_i$, $i=1,\cdots, 7$, are the coefficients. We apply the condition (\ref{genusonefib ; 4.1}). For the pair $(y,z)$ in the base 3-fold $\P^1\times\P^1\times\P^1$, the condition (\ref{genusonefib ; 4.1}) requires that 
\begin{eqnarray}
(\alpha_1\, x^2+\alpha_2 \, xy+\alpha_3\, xz+\alpha_4\, xw+\alpha_5\, yz+\alpha_6\, yw+\alpha_7\, zw)\cdot yz |_Y & = & \\ \nonumber
(\alpha_1\, x^2yz+\alpha_4\, xyzw) (3x+2y+2z+2w) & = & \\ \nonumber
(2\alpha_1+3\alpha_4)\, x^2yzw & = & 0.
\end{eqnarray}
Therefore, we obtain:
\begin{equation}
2\alpha_1+3\alpha_4=0.
\end{equation}
Similarly, by applying the condition (\ref{genusonefib ; 4.1}) to the pairs $(y,w)$ and $(z,w)$, we obtain:
\begin{eqnarray}
2\alpha_1+3\alpha_3 & = & 0 \\ \nonumber
2\alpha_1+3\alpha_2 & = & 0. 
\end{eqnarray}
Thus, we find that the algebraic 2-cycle (\ref{2-cycle in 4.2}) should be of the following form:
\begin{equation}
(\alpha_1\, x^2-\frac{2}{3}\alpha_1 \, xy-\frac{2}{3}\alpha_1\, xz-\frac{2}{3}\alpha_1\, xw+\alpha_5\, yz+\alpha_6\, yw+\alpha_7\, zw)|_Y.
\end{equation}
A K{\"a}hler form $J$ can be expressed as follows:
\begin{equation}
J=a\,x+b\,y+c\,z+d\,w,
\end{equation}
where coefficients $a,b,c,d$ are strictly positive:
\begin{equation}
a, \, b, \, c, \, d >0.
\end{equation}
By applying the condition (\ref{prim ; 4.1}), we obtain:
\begin{eqnarray}
\label{primitivity condition on coeff in 4.2}
(\alpha_1\, x^2-\frac{2}{3}\alpha_1 \, xy-\frac{2}{3}\alpha_1\, xz-\frac{2}{3}\alpha_1\, xw+\alpha_5\, yz+\alpha_6\, yw+\alpha_7\, zw)( a\,x+b\,y+c\,z+d\,w)|_Y &  & \\ \nonumber
\begin{split}
= \, (\alpha_1\, x^2-\frac{2}{3}\alpha_1 \, xy-\frac{2}{3}\alpha_1\, xz-\frac{2}{3}\alpha_1\, xw+\alpha_5\, yz+\alpha_6\, yw+\alpha_7\, zw) & \\
\cdot ( a\,x+b\,y+c\,z+d\,w)(3x+2y+2z+2w) &
\end{split} & &  \\ \nonumber
\begin{split}
=\, a\, (3\alpha_5-\frac{8}{3}\alpha_1)x^2yz+ a\, (3\alpha_6-\frac{8}{3}\alpha_1)x^2yw+ a\, (3\alpha_7-\frac{8}{3}\alpha_1)x^2zw & \\
+[\, 2a(\alpha_5+\alpha_6+\alpha_7)+b(3\alpha_7-\frac{8}{3}\alpha_1)+c(3\alpha_6-\frac{8}{3}\alpha_1)+d(3\alpha_5-\frac{8}{3}\alpha_1)]\, xyzw
\end{split}
&  & \\ \nonumber
=\,  0. & &  
\end{eqnarray}
Thus, we obtain the following conditions on coefficients: 
\begin{eqnarray}
\label{condition coeff in 4.2}
\alpha_5 & = & \frac{8}{9}\alpha_1 \\ \nonumber
\alpha_6 & = & \frac{8}{9}\alpha_1 \\ \nonumber
\alpha_7 & = & \frac{8}{9}\alpha_1.
\end{eqnarray}
Therefore, the algebraic 2-cycle (\ref{2-cycle in 4.2}) should be of the following form:
\begin{equation}
\label{2-cycle with condition in 4.2}
(\alpha_1\, x^2-\frac{2}{3}\alpha_1 \, xy-\frac{2}{3}\alpha_1\, xz-\frac{2}{3}\alpha_1\, xw+\frac{8}{9}\alpha_1\, yz+\frac{8}{9}\alpha_1\, yw+\frac{8}{9}\alpha_1\, zw)|_Y.
\end{equation}
With the conditions (\ref{condition coeff in 4.2}), the equation (\ref{primitivity condition on coeff in 4.2}) reduces to 
\begin{equation}
\frac{16a}{3}\alpha_1\, xyzw=0.
\end{equation}
Thus, we find that that the conditions (\ref{prim ; 4.1}) and (\ref{genusonefib ; 4.1}) require that 
\begin{equation}
\alpha_1=0.
\end{equation}
This means that the algebraic 2-cycle (\ref{2-cycle with condition in 4.2}) vanishes. Thus, we conclude that the conditions (\ref{prim ; 4.1}) and (\ref{genusonefib ; 4.1}) rule out all algebraic 2-cycles obtained as the restrictions of algebraic cycles in the ambient space $\P^2\times\P^1\times\P^1\times\P^1$ to the (3,2,2,2) hypersurface.
\par A similar argument as that stated previously shows that nonintrinsic algebraic 2-cycles of a double cover of $\P^1\times\P^1\times\P^1\times\P^1$ do not yield consistent four-form flux.
\par In Calabi--Yau 4-folds that we constructed, however, it is considerably difficult to explicitly describe intrinsic algebraic 2-cycles. Consequently, it is difficult to compute the self-intersections of intrinsic algebraic 2-cycles in constructed Calabi--Yau 4-folds, and owing to this, it is difficult to determine whether the tadpole can be cancelled using intrinsic algebraic 2-cycles. We do not discuss whether a consistent four-form flux exists. In Section \ref{ssec 4.3} below, we compute the Euler characteristics of Calabi--Yau 4-folds, to derive conditions on the self-intersection of four-form flux to cancel the tadpole.  

\subsection{Euler Characteristics and Self-Intersection of Four-Form Flux to Cancel Tadpole}
\label{ssec 4.3}
\subsubsection{Multidegree (3,2,2,2) Hypersurfaces in $\P^2\times\P^1\times\P^1\times\P^1$}
We compute the Euler characteristic of a multidegree (3,2,2,2) hypersurface $Y$ in $\P^2\times \P^1\times \P^1 \times \P^1$. We have the following exact sequence of bundles: 
\begin{equation}
\label{exact1; 4.3.1}
\begin{CD} 
0 @>>> \mathcal{T}_Y @>>> \mathcal{T}_{\P^2\times \P^1\times \P^1 \times \P^1}|_Y @>>> \mathcal{N}_Y @>>> 0. 
\end{CD}
\end{equation}
$\mathcal{T}_Y$ is the tangent bundle of a genus-one fibered Calabi--Yau multidegree (3,2,2,2) hypersurface $Y$, and this naturally embeds into the tangent bundle $\mathcal{T}_{\P^2\times \P^1\times \P^1 \times \P^1}$ of the ambient space $\P^2\times \P^1\times \P^1 \times \P^1$. $|_Y$ means the restriction to $Y$. $\mathcal{N}_Y$ is the resultant normal bundle. We have 
\begin{equation}
\mathcal{N}_Y\cong \mathcal{O}(3,2,2,2).
\end{equation}
\par From the exact sequence (\ref{exact1; 4.3.1}), we obtain
\begin{equation}
\label{quot Chern ; 4.3.1}
c(\mathcal{T}_Y)=\frac{c(\mathcal{T}_{\P^2\times \P^1\times \P^1 \times \P^1})|_Y}{c(\mathcal{N}_Y)}.
\end{equation}
We have 
\begin{equation}
\label{Chern1 ; 4.3.1}
c(\mathcal{T}_{\P^2\times \P^1\times \P^1 \times \P^1})|_Y=(1+3x+3x^2)(1+2y)(1+2z)(1+2w)|_Y,
\end{equation}
and 
\begin{equation}
\label{Chern2 ; 4.3.1}
c(\mathcal{N}_Y)=1+3x+2y+2z+2w.
\end{equation}
From equations (\ref{quot Chern ; 4.3.1}), (\ref{Chern1 ; 4.3.1}), and (\ref{Chern2 ; 4.3.1}), we can compute $c(\mathcal{T}_Y)$. The top Chern class of $c(\mathcal{T}_Y)$ gives the Euler characteristic of (3,2,2,2) Calabi--Yau hypersurface $Y$. Therefore, we find that 
\begin{equation}
\label{Euler1 ; 4.3.1}
\chi(Y)=1584,
\end{equation}
and 
\begin{equation}
\frac{\chi(Y)}{24}=66.
\end{equation} 
\par We also obtain the second Chern class $c_2(Y)$ from (\ref{quot Chern ; 4.3.1}): 
\begin{equation}
c_2(Y)=(3x^2+6xy+6xz+6xw+4yz+4zw+4wy)|_Y.
\end{equation}
From this, we see that the second Chern class $c_2(Y)$ is not even. 
\par From (\ref{Euler1 ; 4.3.1}), we obtain the net number of 3-branes $N_3$ needed to cancel the tadpole as:
\begin{equation}
\begin{split}
N_3 & =\frac{\chi(Y)}{24}-\frac{1}{2}G_4\cdot G_4 \\
    & =66-\frac{1}{2}G_4\cdot G_4.
\end{split} 
\end{equation}
This must be a non-negative integer, and we therefore obtain a numerical bound on the self-intersection of a four-form flux $G_4$:
\begin{equation}
132 \ge G_4\cdot G_4.
\end{equation}
\par Notice that the result (\ref{Euler1 ; 4.3.1}) of the Euler characteristic is valid for both the Fermat-type hypersurface and the hypersurface in Hesse form.

\subsubsection{Double Covers of $\P^1\times\P^1\times\P^1\times\P^1$ Ramified Along a Multidegree (4,4,4,4) 3-fold}
\par We compute the Euler characteristic of double cover $Y$ of $\P^1\times\P^1\times\P^1\times\P^1$ branched along a (4,4,4,4) 3-fold $B$. The Euler characteristic $\chi(Y)$  of a double cover $Y$ is given by 
\begin{equation}
\chi(Y)=2\cdot\chi(\P^1\times\P^1\times\P^1\times\P^1)-\chi(B).
\end{equation}
We have 
\begin{equation}
\chi(\P^1\times\P^1\times\P^1\times\P^1)=2^4=16,
\end{equation}
therefore
\begin{equation}
\chi(Y)=32-\chi(B).
\end{equation}
\par We use the exact sequence: 
\begin{equation}
\label{exact2}
\begin{CD} 
0 @>>> \mathcal{T}_B @>>> \mathcal{T}_{\P^1\times \P^1\times \P^1 \times \P^1}|_B @>>> \mathcal{N}_B @>>> 0 
\end{CD}
\end{equation}
to obtain the equality 
\begin{equation}
\label{quot Chern2 ; 4.3.2}
c(\mathcal{T}_B)=\frac{c(\mathcal{T}_{\P^1\times \P^1\times \P^1 \times \P^1})|_B}{c(\mathcal{N}_B)}.
\end{equation}
\begin{equation}
\mathcal{N}_B\cong \mathcal{O}(4,4,4,4),
\end{equation}
therefore
\begin{equation}
c(\mathcal{N}_B)=1+4x+4y+4z+4w.
\end{equation}
We have
\begin{equation}
c(\mathcal{T}_{\P^1\times \P^1\times \P^1 \times \P^1})|_B=(1+2x)(1+2y)(1+2z)(1+2w)|_B.
\end{equation}
From the equality (\ref{quot Chern2 ; 4.3.2}), we can compute $c(B)$. The top Chern class of $c(B)$ gives the Euler characteristic $\chi(B)$. Therefore, we deduce that 
\begin{equation}
\chi(B)=-3712.
\end{equation} 
We finally obtain the Euler characteristic $\chi(Y)$:
\begin{equation}
\chi(Y)=32-\chi(B)=32-(-3712)=3744.
\end{equation}
This is divisible by 24: 
\begin{equation}
\frac{\chi(Y)}{24}=156.
\end{equation}   
\par The net number of 3-branes $N_3$ needed to cancel the tadpole is
\begin{equation}
\begin{split}
N_3 &= \frac{\chi(Y)}{24}-\frac{1}{2}G_4\cdot G_4 \\
 & =156-\frac{1}{2}G_4\cdot G_4.
 \end{split}
\end{equation} 
$N_3$ must be a non-negative integer, and therefore, a bound on the self-intersection of four-form flux $G_4$ that we obtain is 
\begin{equation}
312\ge G_4\cdot G_4.
\end{equation}

\section{Matter Spectra and Yukawa Couplings}
\label{sec 5}
We discuss matter fields arising on discriminant components and along matter curves in F-theory compactifications on constructed genus-one fibered Calabi--Yau 4-folds. As discussed in \cite{BHV1}, suppose gauge group $G$ on 7-branes breaks to a subgroup $\Gamma$ such that 
\begin{equation}
\Gamma\times H\subset G
\end{equation}
is maximal. This corresponds to the deformation of singularity associated with gauge group $G$, and consequently, matter fields arise on 7-branes \cite{KV}. When $\Gamma\times H$ has a representation ($\tau, T$), matter fields arise in representation $\tau$ of $\Gamma$, and its generation is given by \cite{BHV1} 
\begin{equation}
n_{\tau}-n_{\tau^*}=-\int_S c_1(S)c_1(\mathcal{T}).
\end{equation}
$S$ denotes a component of the discriminant locus on which 7-branes are wrapped, and $\mathcal{T}$ denotes a bundle transforming in representation $T$ of $H$. We consider the case in which $H$ is $U(1)$. Let $\mathcal{L}$ be a supersymmetric line bundle on component $S$.
\par We discuss matter contents in F-theory compactifications on families of (3,2,2,2) hypersurfaces in Hesse form and double covers of $\P^1\times\P^1\times\P^1\times\P^1$ branched along a multidegree (4,4,4,4) 3-fold below. We focus on specific discriminant components whose forms are isomorphic to $\P^1\times\P^1$. Supersymmetric line bundles on these components are isomorphic to $\mathcal{O}(a,b)$ for some integers $a$ and $b$, $a,b\in\Z$; for line bundles $\mathcal{O}(a,b)$ to be supersymmetric, the integers $a$ and $b$ are subject to the condition $ab<0$ \cite{BHV1}. 
\par As discussed in \cite{BHV1}, Yukawa couplings arise from the following three cases: 
\begin{itemize}
\item interaction of three matter fields on a bulk component
\item interaction of a field on a bulk component and two matter fields localized along a matter curve, and \item triple intersection of three matter curves meeting in a point 
\end{itemize}
Components we consider below have forms isomorphic to $\P^1\times\P^1$, which is a Hirzebruch surface. Therefore, Yukawa coupling does not arise from the first case \cite{BHV1}. We consider Yukawa couplings arising from the second case. 
\par As stated in Section \ref{sec 4}, the existence of a consistent four-form flux is undetermined for Calabi--Yau genus-one fibrations constructed in this note. We can only say that matter contents and Yukawa couplings that we obtain below {\it could} arise. 

\subsection{Matter Spectra for (3,2,2,2) Hypersurfaces in Hesse Form}
We compute matter spectra in F-theory compactifications on (3,2,2,2) hypersurfaces in Hesse Form. We focus on component $A_1$, and we consider the extreme case in which all six components $\{A_i\}_{i=1}^6$ are coincident. $A_1$ is abbreviated to $A$ below. In this case, singular fibers on the bulk $A$ have type $I_6$, and the $SU(6)$ gauge group arises on the 7-branes wrapped on $A$. The form of $A$ is isomorphic to $\P^1\times\P^1$.
\par When $SU(6)$ breaks to $SU(5)$ with
\begin{equation}
SU(6)\supset SU(5)\times U(1),
\label{SU(6) enhancement}
\end{equation}
the adjoint {\bf 35} of $SU(6)$ decomposes as \cite{Sla}:
\begin{equation}
{\bf 35}={\bf 24}_0+{\bf 5}_6+\overline{\bf 5}_{-6}+{\bf 1}_0.
\end{equation}   
Therefore, matter fields {\bf 5} (could) arise on the bulk $A$. The generation of matter fields {\bf 5} on the bulk $A$ is given by:
\begin{equation}
n_{\bf 5}-n_{\overline{\bf 5}}=-\int_Ac_1(A)c_1(\mathcal{L}^{6})=-12(a+b).
\end{equation}
\par $A\cap B_i=\Sigma_1$, $i=1,2,3,4$, and therefore, the bulk $A$ contains four matter curves $\Sigma_1$, which are genus-one curves. When the supersymmetric line bundle $\mathcal{L}$ is turned on, {\bf 20} of $SU(6)$ along matter curve $\Sigma_1$ decomposes as 
\begin{equation}
{\bf 20}={\bf 10}_{-3}+\overline{\bf 10}_{3}.
\end{equation}
Therefore, the mater fields {\bf 10} could localize along a matter curve $\Sigma_1$. 
\par Since matter curve $A\cap B_i=\Sigma_1$ is a bidegree (2,2) curve in $\P^1\times\P^1$, the restriction $\mathcal{L}_{\Sigma_1}$ of the line bundle $\mathcal{L}\cong \mathcal{O}(a,b)$ to matter curve $A\cap B_i=\Sigma_1$ is 
\begin{equation}
\mathcal{L}_{\Sigma_1}\cong \mathcal{O}_{\Sigma_1}(V)
\end{equation}
for some divisor $V$ with deg\,$V=2(a+b)$. We have
\begin{equation}
\begin{split}
n_{\bf 10} & =h^0(K^{1/2}_{\Sigma_1}\otimes \mathcal{L}^{-3}_{\Sigma_1}) \\
           & =h^0(\mathcal{O}_{\Sigma_1}(-3V)).
\end{split}
\end{equation}
Similarly, we have 
\begin{equation}
n_{\overline{\bf 10}}=h^0(3V).
\end{equation}
By the Riemann--Roch theorem, 
\begin{equation}
n_{\bf 10}-n_{\overline{\bf 10}}={\rm deg} (-3V)=-6(a+b).
\end{equation}
\par Therefore, when $a+b>0$ mater fields $\overline{\bf 5}_{-6}$ arise on the bulk $A$, and matter fields $\overline{\bf 10}_3$ localize along matter curve $\Sigma_1$. For this case, Yukawa coupling that arises is 
\begin{equation}
\overline{\bf 5}_{-6}\cdot\overline{\bf 10}_3\cdot\overline{\bf 10}_3.
\label{Yukawa Hesse 1 ; 5}
\end{equation} 
When $a+b<0$, matter fields ${\bf 5}_6$ arise on the bulk $A$, and matter fields ${\bf 10}_{-3}$ localize along matter curve $\Sigma_1$. Yukawa coupling for this case is 
\begin{equation}
{\bf 5}_6\cdot{\bf 10}_{-3}\cdot{\bf 10}_{-3}.
\label{Yukawa Hesse 2 ; 5}
\end{equation}
The results are shown in Table \ref{matter Hesse} below.
\par (3,2,2,2) Calabi--Yau hypersurface in Hesse form has a 3-section, therefore F-theory compactification on it has a discrete $\Z_3$ symmetry \cite{KMOPR, CDKPP, Kdisc}. Thus, massless fields are charged under a discrete $\Z_3$ symmetry; Yukawa coupling has to be invariant under the action of $\Z_3$ \cite{GGK}. We confirm that Yukawa couplings (\ref{Yukawa Hesse 1 ; 5}) and (\ref{Yukawa Hesse 2 ; 5}) indeed satisfy this requirement.

\begingroup
\renewcommand{\arraystretch}{1.5}
\begin{table}[htb]
\begin{flushleft}
  \begin{tabular}{|c|c|c|c|c|c|c|} \hline
Gauge Group & $a+b$ & Matter on $A$ & \# Gen. on $A$ & Matter on $\Sigma_1$ & \# Gen. on $\Sigma_1$ & Yukawa \\ \hline
$SU(6)$ & $>0$ & $\overline{\bf 5}$ & $12(a+b)$ & $\overline{\bf 10}$ & $6(a+b)$ & $\overline{\bf 5}\cdot\overline{\bf 10}\cdot\overline{\bf 10}$ \\ \cline{2-7}
 & $<0$ & {\bf 5} & $-12(a+b)$ & {\bf 10} & $-6(a+b)$ & ${\bf 5}\cdot{\bf 10}\cdot{\bf 10}$ \\ \hline
\end{tabular}
\caption{Potential matter spectra for hypersurface in Hesse form.}
\label{matter Hesse}
\end{flushleft}
\end{table}  
\endgroup 

\subsection{Matter Spectra for Double Covers of $\P^1\times\P^1\times\P^1\times\P^1$ Branched Along a Multidegree (4,4,4,4) 3-fold}
We compute matter spectra in F-theory compactifications on double covers of $\P^1\times\P^1\times\P^1\times\P^1$ branched along a multidegree (4,4,4,4) 3-fold (\ref{doublecov ; 2.2}).
\par When $A_1$ is not coincident with any other $A_i$, $i\ne1$, singular fibers on $A_1$ have type $III$, and $SU(2)$ gauge groups arise on the 7-branes wrapped on $A_1$. For this situation, matter does not arise on the 7-branes wrapped on $A_1$.  
\par When $A_1$ is coincident with another $A_i$, say $A_1=A_2$, $SO(7)$ gauge group arises on the 7-branes wrapped on $A_1$. $A_1$ is abbreviated to $A$. When gauge group $SO(7)$ breaks to $USp(4)$ under 
\begin{equation}
SO(7)\supset USp(4)\times U(1),
\label{SO(7) enhancement}
\end{equation}  
{\bf 21} of $SO(7)$ decomposes as 
\begin{equation}
{\bf 21}={\bf 10}_0+{\bf 5}_{2}+{\bf 5}_{-2}+{\bf 1}_0.
\end{equation}
Therefore, matter fields {\bf 5} (could) arise on the bulk $A$. The generations of {\bf 5} on the bulk $A$ is given by:
\begin{equation}
n_{{\bf 5}_2}-n_{{\bf 5}_{-2}}=-\int_Ac_1(A)c_1(\mathcal{L}^2)=-4(a+b).
\end{equation}
\par $A\cap B_i=\Sigma_9$, $i=1,2$, and therefore, the bulk $A$ contains 2 matter curves \footnote{There are only two matter curves $\Sigma_9$, $A\cap B_1$ and $A\cap B_2$, in component $A$; triple intersection of matter curves in bulk $A$ does not occur for double covers (\ref{doublecov ; 2.2}).} $\Sigma_9$ of genus 9. {\bf 8} of $SO(7)$ decomposes under (\ref{SO(7) enhancement}) as
\begin{equation}
{\bf 8}={\bf 4}_1+{\bf 4}_{-1}.
\end{equation}
Therefore, the matter fields {\bf 4} (could) localize along matter curves $\Sigma_9$. Since $f$ and $g$ are bidegree (4,4) polynomials, the restriction $\mathcal{L}_{\Sigma_9}$ of the line bundle $\mathcal{L}$ to the matter curve $\Sigma_9$ has degree $4(a+b)$. 
The degree of the canonical bundle $K_{\Sigma_9}$ is $2g-2=16$. Let $W$ be the divisor associated with the line bundle $K^{1/2}_{\Sigma_9}\otimes \mathcal{L}_{\Sigma_9}$, so that $\mathcal{O}_{\Sigma_9}(W)=K^{1/2}_{\Sigma_9}\otimes \mathcal{L}_{\Sigma_9}$. The degree of $W$ is $8+4(a+b)$. Now, by the Riemann--Roch theorem, 
\begin{equation}
\begin{split}
n_{{\bf 4}_1}-n_{{\bf 4}_{-1}} & =h^0(W)-h^0(K_{\Sigma_9}-W) \\
& ={\rm deg} \, W+1-9 \\
& =4(a+b).
\end{split}
\end{equation}
Therefore, we have 
\begin{equation}
n_{{\bf 5}_2}-n_{{\bf 5}_{-2}}=-(n_{{\bf 4}_1}-n_{{\bf 4}_{-1}}).
\end{equation}
When $a+b>0$, matter fields ${\bf 5}_{-2}$ arise on the bulk $A$, and matter fields ${{\bf 4}_1}$ localize along matter curves $\Sigma_9$. Yukawa coupling that arises is 
\begin{equation}
{{\bf 5}_{-2}}\cdot{\bf 4}_1\cdot{\bf 4}_1.
\label{Yukawa doublecov 1 ; 5}
\end{equation} 
When $a+b<0$, matters ${\bf 5}_2$ arise on the bulk $A$, and matter fields ${\bf 4}_{-1}$ localise along matter curves $\Sigma_9$. Yukawa coupling for this case is 
\begin{equation}
{\bf 5}_2\cdot {{\bf 4}_{-1}}\cdot {{\bf 4}_{-1}}.
\label{Yukawa doublecov 2 ; 5}
\end{equation}
\par Next, we consider the case in which component $A_1$ is coincident with two other components. Then, singular fiber on $A_1$ are enhanced to type $III^*$, and $E_7$ gauge group arises on the 7-branes wrapped on $A_1$. We again abbreviate component $A_1$ to $A$. When $E_7$ breaks to $E_6$ under 
\begin{equation}
\label{E7 enhancement}
E_7\supset E_6\times U(1),
\end{equation}
{\bf 133} of $E_7$ decomposes as 
\begin{equation}
{\bf 133}={\bf 78}_0+{\bf 27}_{2}+\overline{\bf 27}_{-2}+{\bf 1}_0.
\end{equation}
Therefore, matter fields {\bf 27} (could) arise on component $A$. The generations of {\bf 27} on the bulk $A$ is given by:
\begin{equation}
n_{\bf 27}-n_{\overline{\bf 27}}=-\int_Ac_1(A)c_1(\mathcal{L}^{2})=-4(a+b).
\end{equation}
\par Bulk $A$ contains two matter curves $\Sigma_9$ of genus 9, $A\cap B_i=\Sigma_9$, $i=1,2$. {\bf 56} of $E_7$ decomposes under (\ref{E7 enhancement}) as
\begin{equation}
{\bf 56}={\bf 27}_{-1}+\overline{\bf 27}_{1}+{\bf 1}_3+{\bf 1}_{-3}.
\end{equation}
Therefore, matter fields {\bf 27} localize along the matter curves $\Sigma_9$. The restriction $\mathcal{L}_{\Sigma_9}$ of the line bundle $\mathcal{L}$ to matter curve $\Sigma_9$ has degree $4(a+b)$. Let $W$ be the divisor associated with the line bundle $K^{1/2}_{\Sigma_9}\otimes \mathcal{L}^{-1}_{\Sigma_9}$, so that $\mathcal{O}_{\Sigma_9}(W)=K^{1/2}_{\Sigma_9}\otimes \mathcal{L}^{-1}_{\Sigma_9}$. By applying the Riemann--Roch theorem, we find that the generation of {\bf 27} along matter curve $\Sigma_9$ is given by:
\begin{equation}
\begin{split}
n_{\bf 27}-n_{\overline{\bf 27}} & =h^0(W)-h^0(K_{\Sigma_9}-W) \\
& =-4(a+b).
\end{split}
\end{equation}
When $a+b>0$, matter fields $\overline{\bf 27}$ arise on the bulk $A$, and along matter curves $\Sigma_9$. Yukawa coupling that arises is 
\begin{equation}
\overline{\bf 27}_{-2}\cdot\overline{\bf 27}_1\cdot\overline{\bf 27}_1.
\label{Yukawa doublecov 3 ; 5}
\end{equation}
When $a+b<0$, matter fields {\bf 27} arise on the bulk $A$, and along matter curves $\Sigma_9$. Yukawa coupling for this case is 
\begin{equation}
{\bf 27}_2\cdot{\bf 27}_{-1}\cdot{\bf 27}_{-1}.
\label{Yukawa doublecov 4 ; 5}
\end{equation}
\par Double cover (\ref{doublecov ; 2.2}) has a bisection, and F-theory compactification on it has a discrete $\Z_2$ symmetry \cite{MTsection, Kdisc}. Massless fields are charged under a discrete $\Z_2$ symmetry, and Yukawa coupling has to be invariant under the $\Z_2$ action. We confirm that Yukawa couplings (\ref{Yukawa doublecov 1 ; 5}), (\ref{Yukawa doublecov 2 ; 5}), (\ref{Yukawa doublecov 3 ; 5}), (\ref{Yukawa doublecov 4 ; 5}) satisfy this requirement.
\par The results are shown in Table \ref{matter double cover} below. 

\begingroup
\renewcommand{\arraystretch}{1.5}
\begin{table}[htb]
\begin{flushleft}
  \begin{tabular}{|c|c|c|c|c|c|c|} \hline
Gauge Group & $a+b$ & Matter on $A$ & \# Gen. on $A$ & Matter on $\Sigma_9$ & \# Gen. on $\Sigma_9$ & Yukawa \\ \hline
$E_7$ & $>0$ & $\overline{\bf 27}$ & $4(a+b)$ & $\overline{\bf 27}$ & $4(a+b)$ & $\overline{\bf 27}\cdot\overline{\bf 27}\cdot\overline{\bf 27}$ \\ \cline{2-7}
 & $<0$ & {\bf 27} & $-4(a+b)$ & {\bf 27} & $-4(a+b)$ & ${\bf 27}\cdot{\bf 27}\cdot{\bf 27}$ \\ \hline
$SO(7)$ & $>0$ & {\bf 5} & $4(a+b)$ & {\bf 4} & $4(a+b)$ & ${\bf 5}\cdot{\bf 4}\cdot{\bf 4}$ \\ \cline{2-7}   
 & $<0$ & {\bf 5} & $-4(a+b)$ & {\bf 4} & $-4(a+b)$ & ${\bf 5}\cdot {\bf 4}\cdot {\bf 4}$ \\ \hline
\end{tabular}
\caption{Potential matter spectra for double cover of $\P^1\times\P^1\times\P^1\times\P^1$ (\ref{doublecov ; 2.2}).}
\label{matter double cover}
\end{flushleft}
\end{table}  
\endgroup

\section{Conclusions}
\label{sec 6}
We considered (3,2,2,2) hypersurfaces in $\P^2\times\P^1\times\P^1\times\P^1$, and double covers of $\P^1\times\P^1\times\P^1\times\P^1$ ramified over a (4,4,4,4) 3-fold, to construct genus-one fibered Calabi--Yau 4-folds. By considering specific types of equations, we constructed two families of (3,2,2,2) hypersurfaces, namely Fermat-type hypersurfaces and hypersurfaces in Hesse form. For double covers, we considered a family described by specific types of equations:
\begin{equation}
\label{doublecov ; 6}
\tau^2=f\cdot a(t)\cdot x^4+g\cdot b(t).
\end{equation}
We showed that these three families of genus-one fibered Calabi--Yau 4-folds lack a global section. Genus-one fibers of Fermat-type (3,2,2,2) hypersurfaces and double covers (\ref{doublecov ; 6}) possess complex multiplication of specific orders, 3 and 4, respectively, and these symmetries enabled a detailed study of the gauge theories in F-theory compactifications.
\par We determined the discriminant loci of these families, and we specified the forms of the discriminant components and their intersections. In particular, discriminant components contain matter curves. 
\par $SU(3)$ gauge groups generically arise on 7-branes wrapped on discriminant components in F-theory compactifications on Fermat-type (3,2,2,2) hypersurfaces; when 7-branes coincide, the gauge symmetry is enhanced to $E_6$. Only gauge groups of the form $SU(N)$ arise on 7-branes in F-theory compactifications on (3,2,2,2) hypersurfaces in Hesse form. $SU(2)$ gauge groups generically arise on 7-branes in F-theory compactifications on double covers of $\P^1\times\P^1\times\P^1\times\P^1$ (\ref{doublecov ; 6}). When 7-branes coincide, the $SU(2)$ gauge group is enhanced to $SO(7)$; when more 7-branes coincide, gauge group is enhanced to $E_7$. 
\par We specified the Mordell--Weil groups of Jacobian fibrations of specific Fermat-type hypersurfaces and specific double covers. They are $\Z_3$ and $\Z_2$, such the Mordell--Weil groups have the rank 0, and F-theory compactifications on these specific Calabi--Yau genus-one fibrations do not have $U(1)$ gauge symmetry. 
\par We computed the potential matter spectra and potential Yukawa couplings on specific components. We did not discuss the existence of a consistent four-form flux in this note. We computed the Euler characteristics of Calabi--Yau 4-folds constructed in this note, in order to derive the conditions imposed on four-form fluxes to cancel the tadpole. 

\section*{Acknowledgments}

We would like to thank Shun'ya Mizoguchi and Shigeru Mukai for discussions, and Andreas Kapfer for valuable comments.

\end{document}